\documentclass{aa}
\usepackage{mathtools}
\usepackage{txfonts}
\usepackage{natbib}
\usepackage{graphicx}
\usepackage{stackengine}
\usepackage{epstopdf}
\usepackage{pgf}
\usepackage{longtable}
\usepackage{pdflscape}
\usepackage{geometry}

\usepackage{soul}
\definecolor{lightgreen}{HTML}{B7F774}
\sethlcolor{yellow}

\providecommand{\mc}[1]{\multicolumn{1}{c}{#1}}

\usepackage{hyperref}
\hypersetup{colorlinks,linkcolor={blue},citecolor={blue},urlcolor={red}} 
\providecommand{\abs}[1]{\lvert#1\rvert}

\usepackage{color}
\usepackage{amstext}

\begin{document}

\title{Global millimeter VLBI array survey of ultracompact extragalactic radio sources at 86$\,$GHz}

\author{Dhanya~G.~Nair\inst{1}
       \and
       Andrei~P.~Lobanov\inst{1,2}
       \and
       Thomas~P.~Krichbaum\inst{1}
       \and
       Eduardo~Ros\inst{1}
       \and
       J.~Anton~Zensus\inst{1}
       \and 
       Yuri~Y.~Kovalev\inst{3,4,1}
       \and
       Sang-Sung~Lee\inst{5}
       \and
       Florent~Mertens\inst{6}
       \and
       Yoshiaki~Hagiwara\inst{7}
       \and
       Michael~Bremer\inst{8}
       \and \newline
       Michael~Lindqvist\inst{9}
       \and
       Pablo~de~Vicente\inst{10}
       }
   \institute{Max-Planck-Institut f\"ur Radioastronomie,
             Auf dem H\"ugel 69, 53121 Bonn, Germany
   \and
   Institut f\"ur Experimentalphysik, Universit\"at Hamburg,
   Luruper Chaussee 149, 22761 Hamburg, Germany
   \and
   Astro Space Center of Lebedev Physical Institute, Profsoyuznaya 84/32, 117997 Moscow, Russia
   \and
   Moscow Institute of Physics and Technology, Dolgoprudny, Institutsky per., 9, Moscow region, 141700, Russia
   \and
   Korea Astronomy and Space Science Institute, Daedeokdae-ro 776, Yuseong-gu, Daejeon 34055, Republic of Korea
   \and
   SRON, Kapteyn Astronomical Institute,  Landleven 12, 9747 AD Groningen, Netherlands
   \and
   5-28-20 Hakusan, Bunkyo-ku, Natural Science Laboratory, Toyo University, Tokyo, Japan
   \and
   Institut de Radio Astronomie Millim\'etrique (IRAM), 300 rue de la Piscine, 38406 Saint Martin d'H\`eres, France
   \and
   Department of Space, Earth and Environment, Onsala Space Observatory, Sverige, Observatoriev\"agen 90, Onsala, Sweden
   \and
   Observatorio Astron\'omico Nacional, Observatorio de Yebes, Cerro de la Palera S/N, 19141 Yebes, Spain
 }

\authorrunning{Dhanya~G.~Nair et al.}
\titlerunning{GMVA survey of extragalactic radio sources at 86$\,$GHz} 

  \date{}
 
  \abstract
  % context heading (optional), leave it empty if necessary  
  {Very long baseline interferometry (VLBI) observations at 86$\,$GHz
    (wavelength, $\lambda = 3$\,mm) reach a resolution of about
    50 $\mu$as, probing the collimation and acceleration 
    regions of relativistic outflows in active
    galactic nuclei (AGN). The physical conditions in these regions can be
    studied by performing 86\,GHz VLBI surveys of 
    representative samples of compact extragalactic radio sources.}
  % aims heading (mandatory)
  {To extend the statistical studies of compact extragalactic jets, a
    large global 86\,GHz VLBI survey of 162 compact radio sources was
    conducted in 2010--2011 using the Global Millimeter VLBI Array
    (GMVA).}
  % methods heading (mandatory)
  { The survey observations were made in a snapshot mode, with up to five scans per target 
  spread over a range of hour angles in order to optimize the visibility coverage. The 
  survey data attained a typical baseline sensitivity of 0.1 Jy and a typical
    image sensitivity of 5 mJy/beam, providing successful detections
    and images for all of the survey targets. For 138 objects, the survey provides the first ever VLBI images made at 86 GHz. 
    Gaussian model fitting of the visibility data was applied to represent 
    the structure of the observed sources and to estimate the flux densities and sizes
    of distinct emitting regions (components) in their jets. These estimates were
    used for calculating the brightness temperature
    ($T_\mathrm{b}$) at the jet base (core) and
    in one or more moving regions (jet components) downstream from the core. 
    These model-fit-based estimates of $T_\mathrm{b}$ were compared
    to the estimates of brightness temperature limits made directly
    from the visibility data, demonstrating a good agreement between
    the two methods.}
  % results heading (mandatory)
  { The apparent brightness temperature estimates for the jet cores in our sample 
     range from $2.5 \times 10^{9}$\,K to $ 1.3\times 10^{12}$\,K, 
     with the mean value of $1.8 \times 10^{11} $\,K.
     The apparent brightness temperature estimates for the inner jet components in our sample
      range from $7.0 \times 10^{7}$\,K to $4.0 \times 10^{11}$\,K.
     A simple population model with a
    single intrinsic value of brightness temperature, $T_\mathrm{0}$,
    is applied to reproduce the observed distribution. It yields
    $T_\mathrm{0} = (3.77^{+0.10}_{-0.14}) \times 10^{11}$\,K for the jet
    cores, implying that the inverse Compton losses dominate the
    emission. In the nearest jet components, $T_\mathrm{0} = (1.42^{+0.16}_{-0.19})
    \times 10^{11}$\,K is found, which is slightly higher than the equipartition limit 
    of $\sim5\times 10^{10}$\,K expected for these jet regions. For objects with sufficient
    structural detail detected, the adiabatic energy losses are shown
    to dominate the observed changes of brightness temperature along
    the jet.}
  % conclusions heading (optional), leave it empty if necessary 
   {}

   \maketitle

\section{Introduction}

VLBI (Very long baseline interferometry) observations at 86\,GHz (wavelength, $\lambda=3$\,mm) enable
detailed studies to be made of compact radio sources at a resolution
of $\sim $ (40 --100) $\mu$as. This resolution corresponds to linear
scales as small as 10$^3$--10$^4$ Schwarzschild radii and uncovers the
structure of the jet regions where acceleration and collimation of the
flow takes place \citep{vlahakis2004,lee2008,lee2016,asada2014,boccardi2016,mertens2016}.

To date, five 86\,GHz VLBI surveys have been conducted \citep[][see
Table~\ref{tb:surveys}]{beasley1997,lonsdale1998,rantakyro1998,lobanov2000,lee2008},
with the total number of objects imaged reaching just over a
hundred. No complete sample of objects imaged at 86\,GHz has been
established so far. Recent works \citep[{e.g.},][]{homan2006,cohen2007,lister2016} have demonstrated that
high-resolution studies of complete (or nearly complete) samples of
compact jets yield a wealth of information about the intrinsic
properties of compact extragalactic flows.

Measuring brightness temperature in a statistically viable sample
enables the performance of detailed investigations of the physical conditions
in this region. The distribution of observed brightness temperatures,
$T_\mathrm{ b}$, derived at 86\,GHz can be combined with the
$T_\mathrm{ b}$ distributions measured at lower frequencies
\citep[{e.g.},][]{kovalev2005}. This can help to constrain the
bulk Lorentz factor, $\Gamma_\mathrm{j}$, and the intrinsic brightness
temperature, $T_\mathrm{0}$, of the jet plasma, using different types
of population models of relativistic jets
\citep{vermeulen1994,lobanov2000,lister2003,homan2006}. Changes of
$T_\mathrm{0}$ in the compact jets with frequency can be used to
distinguish between the emission coming from accelerating or
decelerating plasma and from electron-positron or electron-proton
plasma.  Theoretical models predict $T_\mathrm{0} \propto
\nu^{\epsilon}$, with $\epsilon \approx 2.8$, below a critical
frequency $\nu_\mathrm{ break}$ at which energy losses begin to
dominate the emission \citep{marscher1995}. Above $\nu_\mathrm{break}$, 
$\epsilon$ can vary from $-$1 to $+$1, depending on the jet
composition and dynamics. By measuring this break and the power-law
slopes above and below, it would be possible to distinguish between
different physical situations in the compact jets.

Previous studies \citep{lobanov2000,lee2008}
indicate that the value of $\nu_\mathrm{break}$ is likely to be below
86\,GHz. Indeed, a compilation of brightness temperatures measured at
2, 8, 15, and 86\,GHz \citep{lee2008} indicates that
brightness temperatures measured at 86\,GHz are systematically lower
and $\nu_\mathrm{break}$ can be as low as 20\,GHz. This
needs to be confirmed on a complete sample observed at 86\,GHz. 
If $T_\mathrm{0}$ starts to decrease at 86$\,$GHz, there will be only a few
sources suitable for VLBI $> 230$\,GHz and higher
frequencies. Such a decrease of $T_\mathrm{0}$ will also provide a strong
argument in favour of the decelerating jet model or particle-cascade
models as discussed by \cite{marscher1995}. In view of these arguments, it
is important to undertake a dedicated 86$\,$GHz VLBI study of a larger
complete sample of extragalactic radio sources.

In this paper, we present results from a large global VLBI survey of
compact radio sources carried out in 2010--2011 with the Global
Millimeter VLBI Array (GMVA)\footnote{http://www3.mpifr-bonn.mpg.de/div/vlbi/globalmm/.}.
This survey has provided images of 162 unique radio sources,
increasing the total number of sources ever imaged with VLBI at 86$\,$GHz
by a factor of 1.5. The combined database resulting from this survey 
and \cite{lee2008} comprises 256 sources.
This information provides a basis for investigations of
the collimation and acceleration of
relativistic flows and probing the physical conditions in the vicinity of
supermassive black holes.

The survey data reach a typical baseline sensitivity of 0.1\,Jy and a
typical image sensitivity of 5\,mJy/beam.  A total of 162 unique compact radio
sources have been observed in this survey and all the sources are detected and imaged.  
With the present survey, the overall sample of compact radio sources imaged with VLBI at
86\,GHz is representative down to $\sim 0.5$\,Jy for J2000 declinations
of $\delta\ge 15^{\circ}$.

Section~\ref{sc:obs} describes the source selection and the survey
observations. In Section~\ref{sc:data}, we describe the data
processing, amplitude and phase calibration, imaging, model fitting
procedures and a method for estimating errors of the model-fit
parameters. Section~\ref{sc:results} describes the images and derived
parameters of the target sources. Examples of images of four selected
sources obtained from the survey data are presented in
Section~\ref{sc:images} (the complete set of images of all of the
target sources is presented in Appendix available
electronically). Brightness temperatures of the survey sources are
derived and discussed in Section~\ref{sc:tb}. Section~\ref{sc:model}
describes population modelling of the brightness temperature
distribution observed at the base (VLBI core) of the jet and in the
innermost moving jet components. Evolution of the observed brightness
temperature along the jet is studied in Section~\ref{sc:evol} for the
target sources with sufficient extended emission detected.

\section{GMVA survey of compact AGN}
\label{sc:obs}

Dedicated VLBI observations at 86\,GHz are made with the GMVA and with
the Very Long Baseline Array (VLBA)\footnote{Very Long Baseline Array of the National
Radioastronomy Observatory, Socorro, NM; https://www.lbo.us/.}
working in a stand-alone mode (VLBA also takes part in GMVA
observations). The GMVA has been operating since 2002, superseding
the operations of the Coordinated Millimeter VLBI Array
\citep[CMVA;][]{rogers1995} and earlier ad hoc arrangements employed
since the early 1980s \citep{readhead1983}. The GMVA carries out
regular, coordinated observations at 86 GHz, providing good quality
images with a typical angular resolution of $\sim (50$--$70)\,\mu$as.

The array comprises up to 16 telescopes located in Europe, the USA and
Korea operating at a frequency of 86\,GHz. The following telescopes
took part in the GMVA observations for this survey in 2010 and 2011:
eight VLBA antennae equipped with 3\,mm receivers, the IRAM 
(Institut de Radio Astronomie Millim\'etrique) 30\,m
telescope on Pico Veleta (Spain), the phased six-element IRAM
interferometer on Plateau de Bure (France), the MPIfR (Max-Planck-Institut f\"ur Radioastronomie) 
100\,m radio telescope in Effelsberg (Germany), the OSO (Onsala Space Observatory) 20\,m radio telescope at
Onsala (Sweden), the 14\,m telescope in Mets\"ahovi (Finland), and the
OAN (Observatorio Astron\'omico Nacional) 40\,m telescope in Yebes (Spain). 

\subsection{Source selection}

The prime aims of the survey were to establish a complete
sample of compact radio sources imaged with VLBI at 86\,GHz and to
study their morphology and polarization, to study the
distribution of brightness temperatures, to investigate
collimation and 

\begin{figure}[ht!]
\centering
\includegraphics[width=0.5\textwidth,trim={1.0cm 0 1.55cm 1.25cm},clip=true]{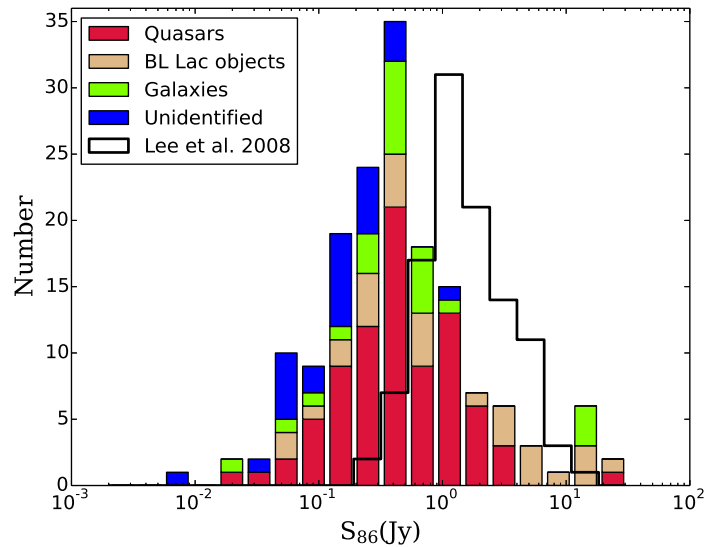}
\caption{Distribution of the total single dish flux
  density of the sources, measured at 86\,GHz at Pico Veleta or Plateau de Bure during the
  observations, $S_{86}$ , broken down according to different
  host galaxy types and compared to the respective distribution for
  the sources from the sample of \cite{lee2008}. The present survey
  provides a nearly twofold increase in the number of objects imaged
  with VLBI at 3\,mm.}
\label{fig:S86}
\end{figure}

\noindent acceleration of relativistic flows and to probe
physical conditions in the vicinity of supermassive black holes.
To meet these aims, the survey target source list has been compiled from the
MOJAVE (Monitoring of Jets in Active Galactic Nuclei with VLBA Experiments) 
sample \citep{kellermann2004,kovalev2005,lister2005,lister2009}, using the following
selection criteria: {\em a)}~15\,GHz correlated flux density, $S_\mathrm{c}$
$\ge 0.5$\,Jy on baselines of $\ge 400$\,M$\lambda$; {\em
b)}~compactness at longest spacings,
$S_\mathrm{c}$/$S_\mathrm{VLBA}$ $\ge 0.4$ where $S_\mathrm{VLBA}$ is
the 15\,GHz total clean flux density; {\em c)}~declination $\delta\ge
15^\circ$. With these selection criteria, a total of 162 unique
sources have been selected, comprising 89 quasars, 26 BL Lac objects,
22 radio galaxies and 25 unidentified sources. Eight bright sources,
3C\,84, OJ\,287, 3C\,273B, 3C\,279, 3C\,345, BL\,Lac, 0716$+$714
and 3C\,454.3 have been added to this list for fringe finding
and calibration purposes. The basic information about the selected
target sources is summarized in Table~\ref{list of sources}.

\begin{table*}[ht!]
\caption{VLBI surveys at 86$\,$GHz}
\label{tb:surveys}
\begin{center}
\begin{tabular}{lcrcrrrrr}\hline\hline
\mc{Survey} & $N_\mathrm{ ant}$ & \mc{$B_\mathrm{rec}$} & $\Delta S$ & \mc{$\Delta I_\mathrm{ m}$} & \mc{$D_\mathrm{ img}$} 
& \mc{$N_\mathrm{ obs}$} & \mc{$N_\mathrm{ det}$} & \mc{$N_\mathrm{ img}$} \\
\mc{(1)} & (2) & \mc{(3)} & (4) & \mc{(5)} & \mc{(6)} & \mc{(7)} & \mc{(8)} &
\mc{(8)} \\ \hline
Beasley et al. (1997)     & 3    &   112    &$\sim 0.5$ & ... & ... &  51  & 12 &  ... \\
Lonsdale et al. (1998)    & 2--5 &   112/224&$\sim 0.7$ & ... & ... &  79  & 14 &  ... \\
Rantakyr\"o et al. (1998) & 6--9 &   128    &$\sim 0.5$ & $\sim 30$ & 70 &  68  & 16 &  12  \\
Lobanov et al. (2000)     & 3--5 &   224    &$\sim 0.4$ & $\sim 20$ & 100 &  28  & 26 &  14  \\ 
Lee et al. (2008)         & 12   &   256    &$\sim 0.3$ &  $10 $ & 200 & 127 & 121 & 109 \\ \cline{2-9}
{\em This survey}    & 13--14    &   512  & $\sim 0.1$ &$\sim 5$ & $>400$ & 162 & 162 & 162 \\ \hline  
\end{tabular}
\end{center}
\footnotesize{ {\bf Columns:} 1~--~Survey ; 2~--~number of participating antennae; 
  3~--~recording bit rate [Mbps]; 4~--~average baseline sensitivity [Jy]; 5 -- average image
  sensitivity [mJy/beam]; 6~--~typical dynamic range of images;
  7~--~number of sources observed; 8~--~number of  sources detected;
  9~--~number of  sources imaged.} 
\end{table*}

\begin{table*}[ht!]
\caption{Log of survey observations}
\label{log of survey observations}
\begin{center}
\begin{tabular}{ccccccccc}\hline\hline
\mc{Part}& \mc{Date} & \mc{$N_\mathrm{obj}$ } &  \mc{Pol.} & \mc{$w_\mathrm{bit}$} &    \mc{BW}  & \mc{$n_\mathrm{ch}$}  &     \mc{$n_\mathrm{bit}$}  &  \mc{Telescopes}    \\ 
\mc{(1)}  & \mc{(2)}  & \mc{(3)}                   & \mc{(4)}       & \mc{(5)}        & \mc{(6)}                & \mc{(7)}     & \mc{(8)}  & \mc{(9)} \\ \hline
A    &  Oct 2010    &    68    & LCP           &   512      & 128 (4IF x 32)       &   32   & 2   & 8 VLBA+(Eb,On,Mh,Pb,Pv)         \\ 
B    &  May 2011    &    46     & LCP          &   512      & 128 (2IF X 64)       &   64   & 2   & 8 VLBA+(Eb,On,Pb,Pv,Mh)          \\ 
C    &  Oct 2011    &    60     & LCP         &   512      &  128 (4IF x 32)       &   32  & 2   & 8 VLBA+(Eb,On,Pb,Pv,Mh,Ys)       \\ \hline
\end{tabular}
\end{center}
\footnotesize{ {\bf Columns:} 1~--~survey epoch; 2~--~observing date;
  3~--~number of objects observed; 4~--~polarization;
  5~--~recording bit rate [Mbps]; 
  6~--~total bandwidth (number of IF bands x IF bandwidth) [MHz]; 7~--~number of frequency channels per IF band; 8~--~correlator sampling [bits];
  9~--~participating telescopes} 
\end{table*}

\begin{figure*}[ht!]
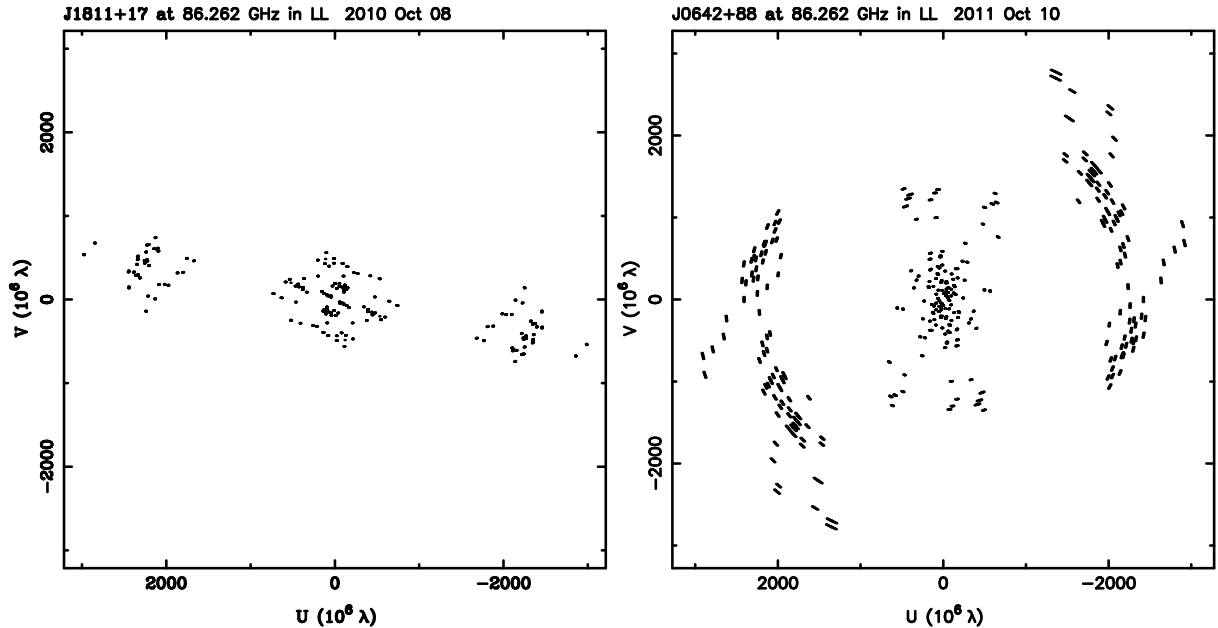

\centerline{\includegraphics[width=0.45\textwidth,angle=-90]{Plots/uvplot_low_dec.eps}
            \includegraphics[width=0.45\textwidth,angle=-90]{Plots/uvplot_high_dec.eps}
}
\caption{Examples of {\em uv}-coverages of the survey observations for a
  low declination source (left; J1811+1704, $\delta=+17^{\circ}$) and a
  high declination source (right; J0642+8811, $\delta = +88^{\circ}$). 
}
\label{fig:uvplots}
\end{figure*}

The distribution of the total single dish flux
of the sources, measured at 86\,GHz at Pico Veleta or Plateau de Bure during the
observations, is shown in Figure~\ref{fig:S86} and compared with the
respective distribution of the source sample observed in
\cite{lee2008}. This comparison shows that our survey observations
probe objects at about one order of magnitude weaker sources and
provide a roughly twofold increase of the total number of objects
imaged with VLBI at 3\,mm wavelength (see Table~\ref{list of sources} for details).

\subsection{Observations}

The survey observations have been made over a total of six days (144
hours), scheduled within three separate GMVA sessions. Up to 14
telescopes took part in the observations (see Table~\ref{log of
  survey observations}). The observations were typically scheduled
with five scans per hour, each of 300 seconds in duration. Gaps of five to ten
minutes were introduced between the scans for antennae pointing at
Effelsberg (Eb) and Pico Veleta (Pv) and
for phasing of the Plateau de Bure (Pb) interferometer. This observing
scheme yielded the total of 720 scans distributed between 174
observing targets (162 unique radio sources), ensuring that each
object was observed with four to five scans distributed over a wide range of
hour angles. Despite the rather modest observing time spent on each
target, the large number of participating antennae ensured
good {\em uv}-coverages for all survey sources down to the lowest
declinations (see Figure~\ref{fig:uvplots}).

The observations were performed at a sampling rate of 512$\,$Mbit/sec
and with a two--bit sampling. There were four intermediate frequencies (IFs) 
in Epoch A and C, and two IFs in Epoch B. The typical baseline sensitivities for a 20--second
integration time are $\approx 0.05\,$Jy on the Pb--Pv baseline,
$\approx 0.1\,$Jy on the Eb--Pv baseline, $\approx 0.2\,$Jy on the
baselines between Eb/Pv and other antennae, and $\approx 0.4$\,Jy on the
baselines between the VLBA antennae. With such baseline sensitivities
and an on-source integration time of about 20 minutes,
the typical point source sensitivity of a survey observation is
$\sim 5\,$mJy/beam, which is sufficient to obtain robust images of
most of the survey sources.

\begin{table*}[ht!]
\caption{Participating telescopes}
\label{participating telescopes}
\begin{center}
\begin{tabular}{lcccccccc}\hline\hline
                &      & $D$ & G & $T_\mathrm{sys}$ & $\eta_\mathrm{A}$ & SEFD & $\Delta_\mathrm{512,20} $ & $\sigma_\mathrm{rms}$ \\
\mc{Name}       & Code & [m] & [K/Jy] & [K] & &  [Jy] & [mJy] & [mJy] \\
\mc{(1)}        & (2)  & (3) & (4)  & (5) & (6) & (7) & (8) & (9)  \\ \hline
Brewster        & Br   & 25  & 0.033   & 110 & 0.22 & 3333.3 & 23.44 & 164.09   \\
Effelsberg      & Eb   & 80$^\dag$  & 0.140  & 140 & 0.08 & 1000 & 12.84  & 89.87 \\
Fort Davis      & Fd   & 25  & 0.039  & 140 & 0.22 & 3589.7 &  24.32  & 170.28   \\
Kitt Peak       & Kp   & 25  & 0.028  & 100 & 0.22 & 3571.4 & 24.26 & 169.85    \\
Los Alamos      & La   & 25  & 0.042  & 120 & 0.22 & 2857.1 & 21.70  & 151.91    \\     
Mets\"ahovi     & Mh   & 14  & 0.017   & 300 & 0.3 & 17647.1 & 53.94 & 377.55   \\
Mauna Kea       & Mk   & 25  & 0.019   & 100 & 0.22 & 5263.2 & 29.45  & 206.18   \\
North Liberty   & Nl   & 25  & 0.022   & 130 & 0.22 & 5909.1 & 31.21  & 218.47    \\
Onsala          & On   & 20  & 0.049   & 250 & 0.43 & 5102 & 29.00   & 203.00   \\
Owens Valley    & Ov   & 25  & 0.035   & 120 & 0.22 & 3428.6 & 23.77  & 166.41   \\
Plateau de Bure & Pb   & 36$^\ddag$  & 0.22 &  180 & 0.7 & 818.2 & 11.61  & 81.29    \\
Pie Town        & Pt   & 25  & 0.044  & 100 & 0.22 & 2272.7 & 19.36  & 135.52   \\ 
Pico Veleta     & Pv   & 30  & 0.153  & 100 & 0.6 & 653.6 & ... & ...   \\ 
Yebes           & Ys   & 40  & 0.09  & 150 & 0.2 & 1666.7 & 16.58  & 116.03   \\ \hline  
\end{tabular}
\end{center}
\footnotesize{ {\bf Columns:} 1~--~telescope name; 2~--~abbreviation for the telescope name;
  3~--~diameter; 4~--~antenna gain; 5~--~zenith system temperature; 6~--~aperture efficiency; 
7~--~zenith SEFD; 8~--~sensitivity on the baseline to Pico Veleta, for a 20 sec fringe fit 
interval and 512 Mbps recording rate; 9~--~7$\sigma$ detection threshold. {\bf Notes:} 
$^\dag$ -- effective diameter at 86\,GHz; $^\ddag$ -- effective diameter in the phased array mode.} 
\end{table*}

\section{Data processing}
\label{sc:data}

\subsection{Correlation and fringe fitting}

The data were correlated at the DiFX correlator \citep{deller+2011} of
the Max-Planck-Institut f\"ur Radioastronomie (MPIfR) at Bonn. After correlation, the
data were loaded into AIPS \citep[Astronomical Image Processing
System;][]{greisen1990}. After applying the correlator model, the residual fringe delays and
rates were determined and corrected for both within the individual
IFs (single-band fringe fitting) and between the IFs (multi-band fringe fitting).  

At the first step of the fringe fitting, manual phasecal correction
was applied by obtaining the single band delay and delay rate
solutions from a scan on a strong source that gives high signal-to-noise ratio (SNR)
of the fringe solutions (an $\mathrm{SNR}\ge 7$ threshold was set) over the typical
coherence time ($\approx 20\,\mathrm{sec}$) for all the antennae. The resulting
fringe solutions were applied to the entire dataset.  After the manual
phasecal correction, antenna-based global fringe fitting \citep{schwab1983,alef1986} 
was performed, setting the solution interval to the full scan length in
order to improve the detection SNR. Pico Veleta was used as the
reference antenna for almost all the data. Whenever Pico Veleta was
not present in the data, Plateau de Bure or Pie Town were used as the
reference antennae. To minimize the chance of false detections, the
data were fringe fitted with a SNR threshold of five and a search window
width of 200 nsec for the fringe delay and 400 mHz for the fringe
rate.

Once the global fringe fit was done, the SNR of the fringe solutions
were inspected for all the sources, and strong sources which give
relatively very high SNR were determined.  Whenever feasible, the
solutions from those strong sources were interpolated to nearby weak
sources, using a procedure similar to that adopted in the similar
earlier VLBI survey observations at 86\,GHz \citep[cf.,][]{lee2008}.
Application of the interpolated solutions has resulted in the detection of amplitude
and phase signals for all of the survey targets.

\subsection{Amplitude calibration}

A priori amplitude calibration was done using the measured values of
antenna gain and system temperature. The weather information from each
station during the observation was used to correct for the atmospheric
opacity.  The initial opacity correction was implemented by setting
the opacity $\tau =0.1$ and fitting for the receiver temperatures,
$T_\mathrm{rec}$.  At the second step, the fitted receiver
temperatures were used as initial values for fitting simultaneously
for $\tau$ and $T_\mathrm{rec}$.

The accuracy and self-consistency of the amplitude calibration was
checked with a procedure developed and used in the earlier 86\,GHz
survey experiments \citep{lobanov2000,lee2008}. The calibrated
visibility data were model fitted for each of the survey targets, using
two-dimensional Gaussian components and allowing for scaling the
individual antenna gains by a constant factor.  The resulting average
gain scale corrections are listed in Table~\ref{antenna gain}. The
scale offsets are within 25\,$\%$ for most of the antennae, except
Mets\"ahovi which had suffered from persistently bad weather during
each of the three observing sessions. The averaged offsets are within
about 3\,$\%$ of the unity implying that there is no significant bias
in the calibration and their r.m.s. (root mean square) is less than 10\,$\%$ for each of
the three observing epochs. Based on this analysis, we conclude that
the a priori amplitude calibration should be accurate to within about
25\,$\%$, providing a sufficient initial calibration accuracy for the
hybrid imaging of the source structure.

\begin{table}[ht!]
\caption{Average antenna gain corrections}
\label{antenna gain}
\begin{center}
\begin{tabular}{lccc}\hline\hline
\mc{Telescope}  & Epoch A & Epoch B & Epoch C \\
\mc{(1)}        & (2)  & (3) & (4)    \\ \hline
Br              &  1.008$\pm$0.406 & 1.054$\pm$ 0.137  &  1.010$\pm$ 0.192   \\
Eb              &  1.103$\pm$0.264 & 1.028$\pm$ 0.224  &  0.925$\pm$ 0.182  \\
Fd              &  0.992$\pm$0.206  & 1.105$\pm$ 0.289  & 0.958$\pm$ 0.162   \\
Kp              &  0.881$\pm$ 0.154 & 0.933$\pm$ 0.104  & 0.931$\pm$ 0.102    \\
La              &  0.819$\pm$ 0.199 & 0.940$\pm$ 0.172  & 0.987$\pm$ 0.133   \\
Mh              &  ...  & ...  &  1.620$\pm$ 0.573 \\
Mk              &  0.919$\pm$ 0.256 & 1.188$\pm$ 0.438  & 0.844$\pm$ 0.199    \\
Nl              &  1.224$\pm$ 0.386 & 1.190$\pm$ 0.296  &  0.798$\pm$ 0.264   \\
On              &  0.934$\pm$ 0.292  & 0.935$\pm$ 0.114  & 0.889$\pm$ 0.147    \\
Ov              &  0.900$\pm$ 0.371 & 0.868$\pm$ 0.148  & 0.939$\pm$ 0.382     \\
Pb              &  1.081$\pm$ 0.146 & 1.087$\pm$ 0.225  & 1.095$\pm$ 0.203    \\
Pt              &  0.841$\pm$ 0.113 & 0.878$\pm$ 0.109  & 1.041$\pm$ 0.098    \\
Pv              &  1.006$\pm$0.116 &  1.088$\pm$ 0.220 &  1.010$\pm$ 0.086   \\
Ys              &  ...  & ...  &   1.009$\pm$0.196   \\    
Array average   &  1.037$\pm$0.084 &0.991$\pm$ 0.074   &  1.027$\pm$0.097   \\    \hline
\end{tabular}
\end{center}
\footnotesize{ {\bf Columns:} 1~--~abbreviation for the participating telescope; 
2,3,4~--~average and r.m.s of antenna gains for observing epochs A, B, and C, respectively.}
\end{table}

\subsection{Hybrid imaging and model fitting}

After the phase and amplitude calibration was applied on the data, the
visibility data were averaged for 10 sec for most of the sources and
in some cases, the data were averaged for 30 sec. The data were then processed in the the Caltech
DIFMAP software package \citep{shepherd1994}, modelling them with
circular Gaussian patterns
\citep[model fitting;][]{fomalont1999} and obtaining hybrid images.
The data were not {\em uv}-tapered and the imaging was performed using
the natural weighting.

\begin{figure}[ht!]
  \centering
\includegraphics[width=0.5\textwidth,trim={1.0cm 0 1.55cm 1.25cm},clip=true]{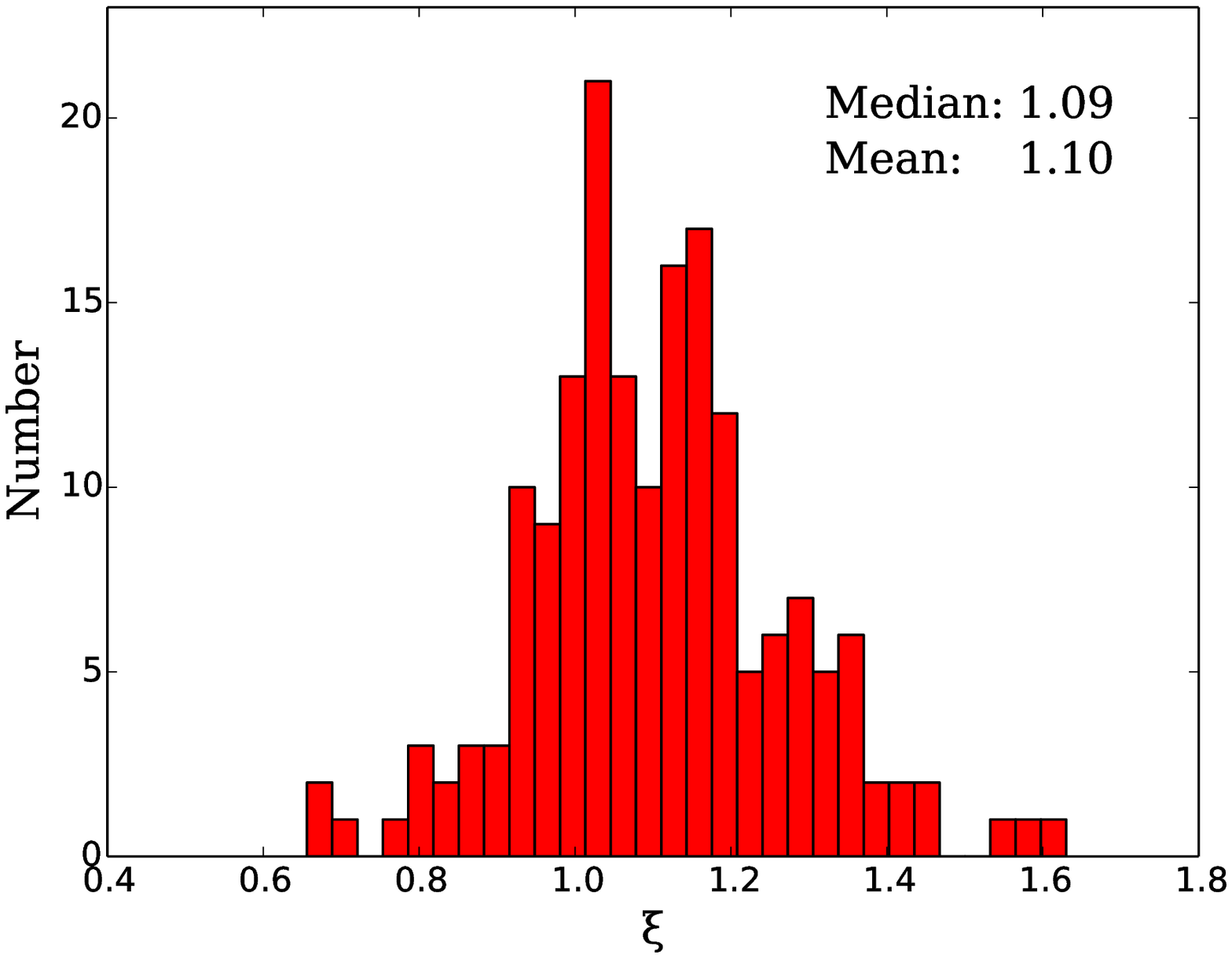}
\caption{Distribution of the noise quality factor $\xi_\mathrm{r}$ for
  the residual images of all the sources in the survey.}
\label{fig:qualityfactor}
\end{figure}

\begin{figure}[h!]
\centering
\includegraphics[width=0.5\textwidth,trim={1.0cm 0 1.55cm 1.25cm},clip=true]{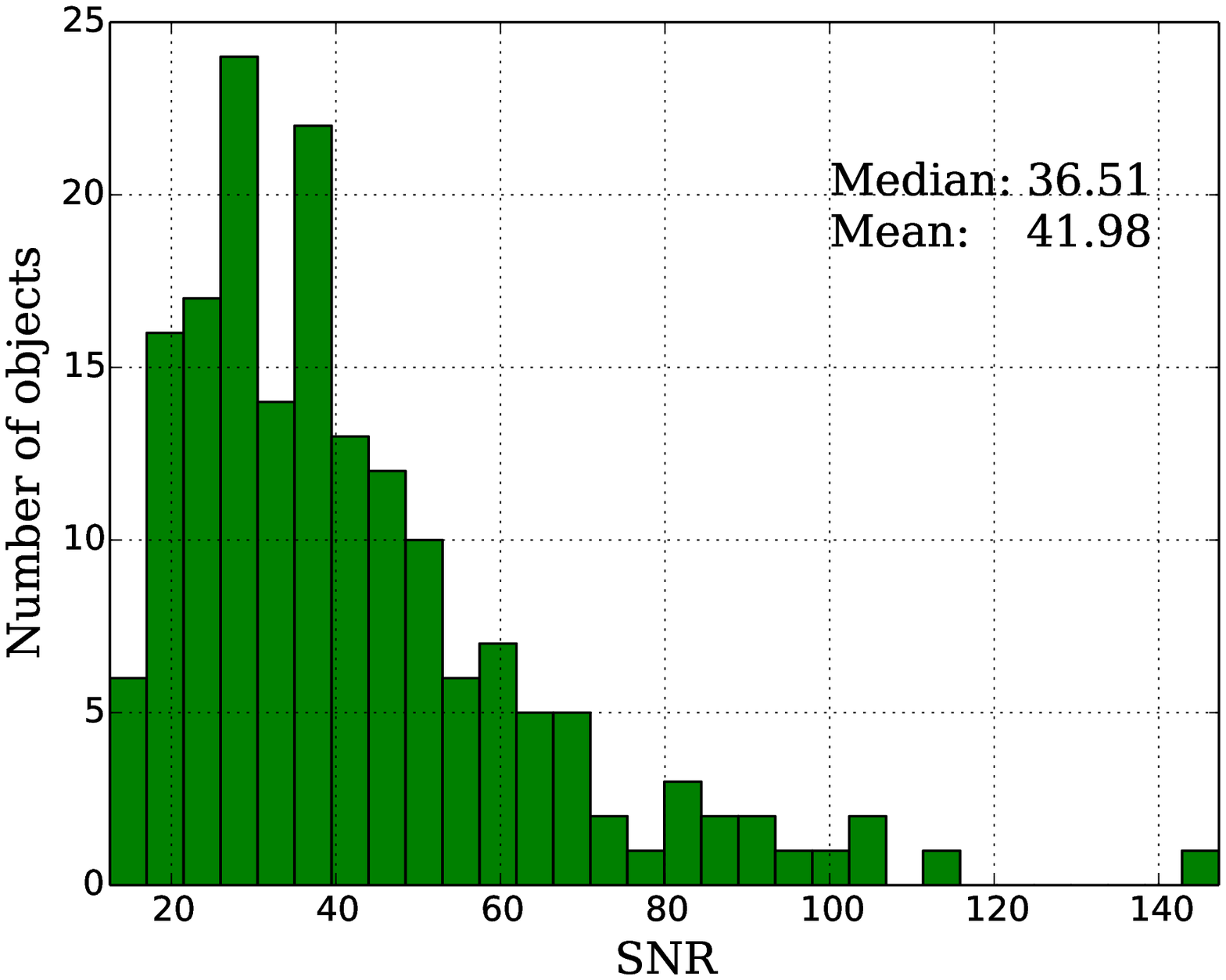}
\caption{Distribution of the imaging signal-to-noise ratios
  (SNR) of all the sources in the survey calculated from the ratio of
  the peak flux density of the map, $S_\mathrm{p}$ , to the r.m.s noise in
  the map, $\sigma_\mathrm{rms}$.}
\label{fig:Tb_SNR}
\end{figure}

The initial model fitting was performed on the calibrated data and was used,
in some cases, to facilitate the hybrid imaging. The final model fitting was done 
on the self-calibrated data resulting from the hybrid imaging.
The total number of Gaussian components used for model fitting a given
source was determined using the $\chi^2$ statistics of the fits. The
final models were obtained when the addition of another Gaussian
component did not provide a statistically significant change of the
$\chi^{2}$ agreement factors \citep{schinzel2012}. 

The hybrid imaging procedure comprised the CLEAN deconvolution
\citep{clark1980} and self calibration
\citep{cornwell1999,cornwell1995}. For most of the objects, the
hybrid imaging procedure was initiated with a point source model.  For
objects with sparse visibility data, the initial Gaussian model fits were used
as the initial models. Only visibility phases were allowed to be
modified during the initial iterations of the hybrid imaging. At the
last step, a single time constant antenna gain correction factor was
applied to the visibility amplitudes (hence not allowing for time
variable antenna gains in order to avoid the imprint of model errors into data). 
The parameters of the final images are listed
in Table~\ref{Image Parameters}, together with the correlated flux
densities measured on short and long baselines.

The quality of the
residual noise in the final images, which ideally should have a
zero-mean Gaussian distribution, was checked by calculating the
expectation value for the maximum absolute flux density $\abs
{S_\mathrm{r,exp}}$ in a residual image \citep{lee2008},
    \begin{equation}
     \abs{S_\mathrm{r,exp}}= \sigma_\mathrm{r} \sqrt{2} \ln \left( \frac{N_\mathrm{pix}}{\sqrt{2\pi\sigma_\mathrm{r}}} \right)^\frac{1}{2}\,,
    \end{equation}   
    where $N_\mathrm{pix}$ is the total number of pixels in the
    image. The quality of the residual noise is given by
    $\xi_\mathrm{r}$ = $S_\mathrm{r}$/$S_\mathrm{r,exp}$ , where
    $S_\mathrm{r}$ is the maximum flux density in the residual image
    and $\sigma_\mathrm{r}$ is the r.m.s noise in the residual image. When the
    residual noise approaches Gaussian noise, $\xi_\mathrm{r}$ tends
    to 1. If $\xi_\mathrm{r}$ $>$ 1, not all the structure has been
    adequately recovered; if $\xi_\mathrm{r}$ $<$ 1, the image model
    has an excessively large number of degrees of freedom
    \citep{lobanov2006}. Figure~\ref{fig:qualityfactor} shows the
    overall distribution of $\xi_\mathrm{r}$ for the residual maps of
    all the sources in this survey. Column 14 in Table~\ref{Image
      Parameters} shows the quality factor, $\xi_\mathrm{r}$ obtained
    for all the 3\,mm images implying that the images adequately
    represent the source structure detected in the visibility
    data. The Gaussianity of the residual noise is also reflected in
    the median and the mean of the $\xi_\mathrm{r}$ distributions,
    which are within 10\,$\%$ of the unity factor.

In order to check the effect of the amplitude (antenna gain)
corrections applied during the final self calibration step, we
compared the visibility amplitudes obtained without and with it. This
was done by comparing the ratios of the visibility amplitudes obtained
without and with the antenna gain correction on short $B_\mathrm{S}$
and long $B_\mathrm{L}$ baselines (see Table~\ref{Image Parameters}). 
The average of the ratios are found to be (1.24$\pm$ 0.18) and (1.01 $\pm$ 0.14) for the short and long baselines,
respectively. These ratios imply that the amplitude self-calibration did not introduce substantial gain corrections, thus further 
indicating the overall good quality of the a priori amplitude calibration of the survey data.

\subsection{Gaussian models of the source structure}

The final self calibrated data were fitted with circular Gaussian
components, using the initial model fits as a starting guess.  The
resulting models were used to obtain the total, $S_\mathrm{tot}$, and
peak flux density, $S_\mathrm{peak}$, the size, $d$, and the
positional offset, (in polar coordinates $r$,$\theta$) of the
component from the brightest region (core) at the base of the jet,
taken to be at the coordinate origin.  The uncertainties of the
model parameters were estimated analytically, based on the SNR of
detection of individual components, following \citep{fomalont1999} and
\citep{schinzel2012}:
\begin{eqnarray}
\label{un1}
\sigma_{\rm peak} = \sigma_{\rm rms}{\left({1+\frac{S_{\rm peak}}{\sigma_{\rm rms}}}\right)}^{1/2}, & 
\sigma_{\rm tot}  = \sigma_{\rm peak}{\left({1+\frac{S_{\rm tot}^2}{S_{\rm peak}^2}}\right)}^{1/2}, \nonumber
\end{eqnarray}
\begin{eqnarray}
\label{un2}
\sigma_{d} = d\frac{\sigma_{\rm peak}}{S_{\rm peak}}, & 
\sigma_{r} = \frac{1}{2}\sigma_{d}, & 
\sigma_{\theta} ={\rm atan}\left(\frac{\sigma_{r}}{r}\right),
\end{eqnarray}
where $\sigma_\mathrm{rms}$ is the r.m.s. noise in the residual image after substraction of the Gaussian model fit.    
To assess whether a given component is extended (resolved), the minimum resolvable size of the component
was also calculated and compared with the size obtained with the model fitting. The minimum resolvable size, $d_\mathrm{min}$ of a Gaussian component is given in \cite{lobanov2005} as
 \begin{equation}
 \label{eqn:dmin}
     d_\mathrm{min}= \frac{2^{\left(1+\beta/2\right)}}{\pi}  \left[ \pi \, a \, b \, \log \left( \frac{\mathrm{SNR}+1}{\mathrm{SNR}} \right) \right] ^{1/2}\,,
    \end{equation}
where $a$ and $b$ are the axes of the restoring beam, SNR is the signal-to-noise ratio, and $\beta$ is the weighing function that is 0 for natural weighting or 
2 for uniform weighting. For components that have size estimates $<d_\mathrm{min}$, the latter
is taken as an upper limit of the size and used for estimating the uncertainties of the other model fit parameters of the respective 
component (see Table~\ref{Model fit parameters}).

\begin{figure*}[ht!]
\centerline{
\includegraphics[width=0.5\textwidth,trim={1.0cm 0 1.55cm 1.25cm},clip=true]{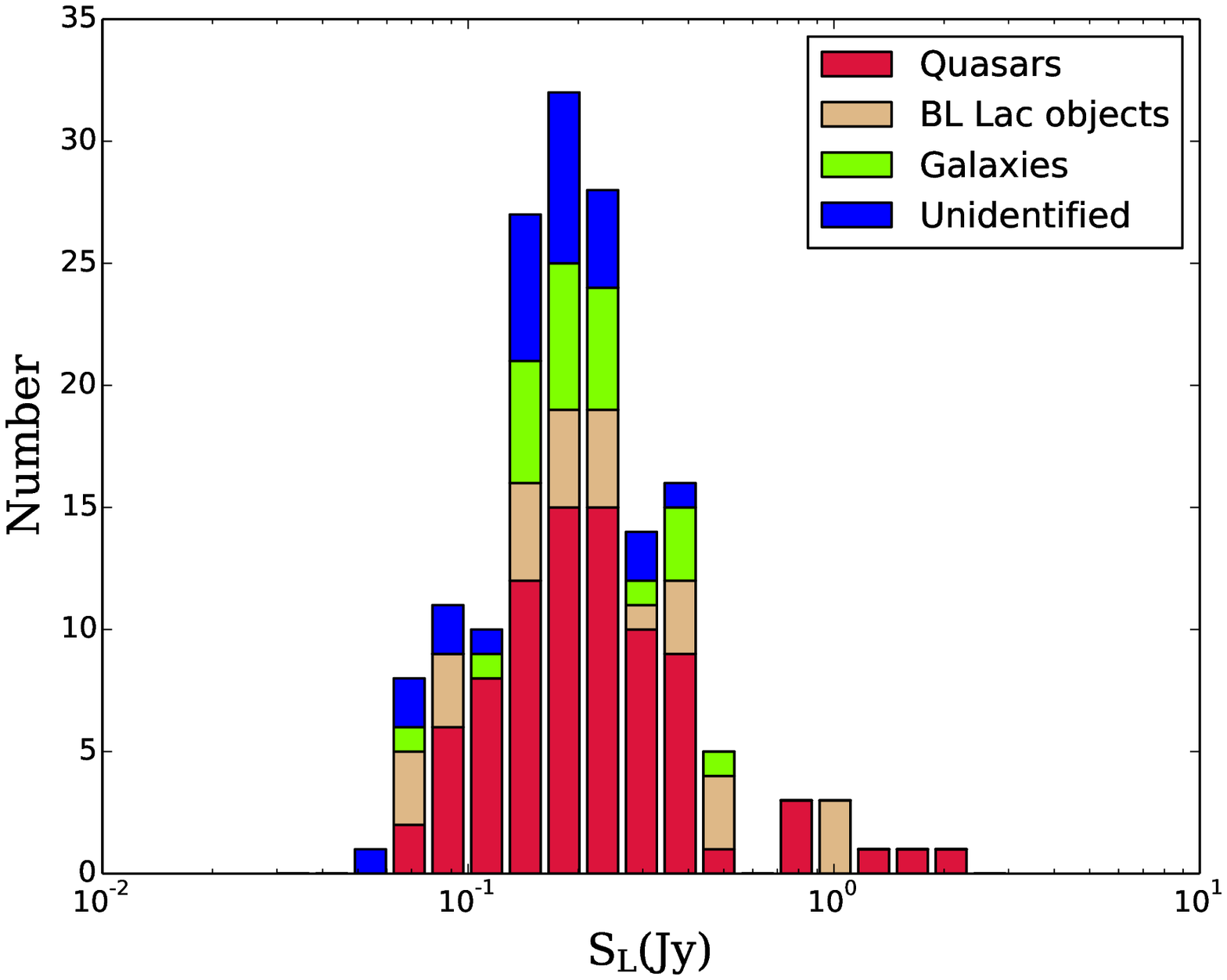}
\includegraphics[width=0.5\textwidth,trim={1.0cm 0 1.55cm 1.25cm},clip=true]{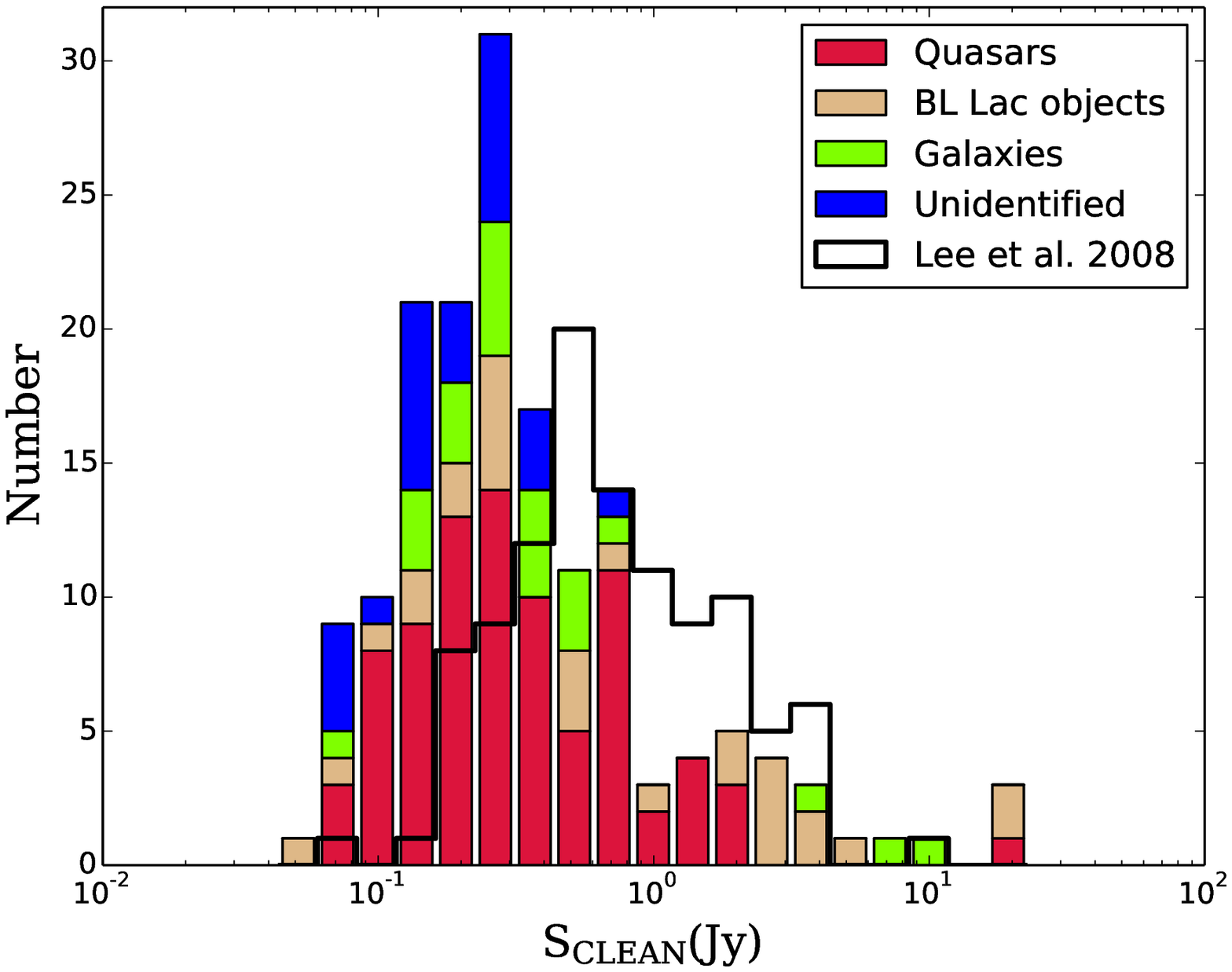}
}
\caption{Distribution of the correlated flux densities corresponding to the longest baselines,
  $S_\mathrm{L}$ (left) and the total clean image flux density, $S_\mathrm{CLEAN}$ of the survey targets broken down according to different
  host galaxy types (right). The distribution of $S_\mathrm{CLEAN}$ for the sources in this survey is also compared with the respective distribution for the
  sources from the sample of \cite{lee2008} on the right panel.}
\label{fig:Sclean_SL}
\end{figure*}

\subsection{Brightness temperature estimates}

We use the total flux density $S_\mathrm{tot}$ and size $d$ of the
model fit components to estimate the brightness temperature,
$T_\mathrm{b} = I_\nu c^2 / 2k_\mathrm{B}\nu$ (with $\nu$, $k_\mathrm{B}$, and $c$ denoting the observing frequency, the Boltzmann
constant, and the light speed, respectively) of the individual emitting
regions in the jets. \par
For a circular Gaussian component, $I_\nu =
(4\,\log 2/2) S_\mathrm{tot}/d^2$, and the respective brightness
temperature can be obtained from
\begin{equation}
\label{eqn:tb}
T_\mathrm{b}\mathrm{[K]} = 1.22 \times 10^{12} \left(\frac{S_\mathrm{tot}}{\mathrm{Jy}}\right) \left(\frac{d}{\mathrm{mas}}\right)^{-2} 
\left(\frac{\nu}{\mathrm{GHz}}\right)^{-2} (1+z) \,.
\end{equation}
The factor $(1+z)$ reflects the effect of the
cosmological redshift $z$ on the observed brightness temperature. For
the sources with unknown redshift, we calculated the brightness
temperature simply in the observer's frame of reference. If the size
of the Gaussian component $d$ is less than $d_\mathrm{min}$, given
by Equation (\ref{eqn:dmin}), the latter is used for estimating the
lower limit on $T_\mathrm{b}$.

In addition to this estimate, we also use visibility-based estimates of brightness temperature \citep{lobanov2015} and calculate the minimum brightness
temperature,
\begin{equation}
T_\mathrm{b,min}\mathrm{[K]}= \, 3.09 \left(\frac{V_\mathrm{q}}
{\mathrm{mJy}}\right) \left(\frac{B_\mathrm{L}}{\mathrm{km}}\right)^{2}\,, 
\label{eq:tbmin}
\end{equation}
and limiting resolved brightness temperature,
\begin{equation}
T_\mathrm{b,lim}\mathrm{[K]} = 1.14 \left(\frac{V_\mathrm{q}+\sigma_\mathrm{q}}{\mathrm{mJy}}\right)\left(\frac{B_\mathrm{L}}{\mathrm{km}}\right)^{2} \left(\ln \frac{V_\mathrm{q}+\sigma_\mathrm{q}}{V_\mathrm{q}}\right)^{-1}\,,
\label{eq:tblim}
\end{equation}
directly from the visibility amplitude, $V_\mathrm{q}$, and its error,
$\sigma_\mathrm{q}$, measured at a given long baseline,
$B_\mathrm{L}$, in the survey data.

\section{Results}
\label{sc:results}

\subsection{Images}
\label{sc:images}

\begin{figure*}[ht!]
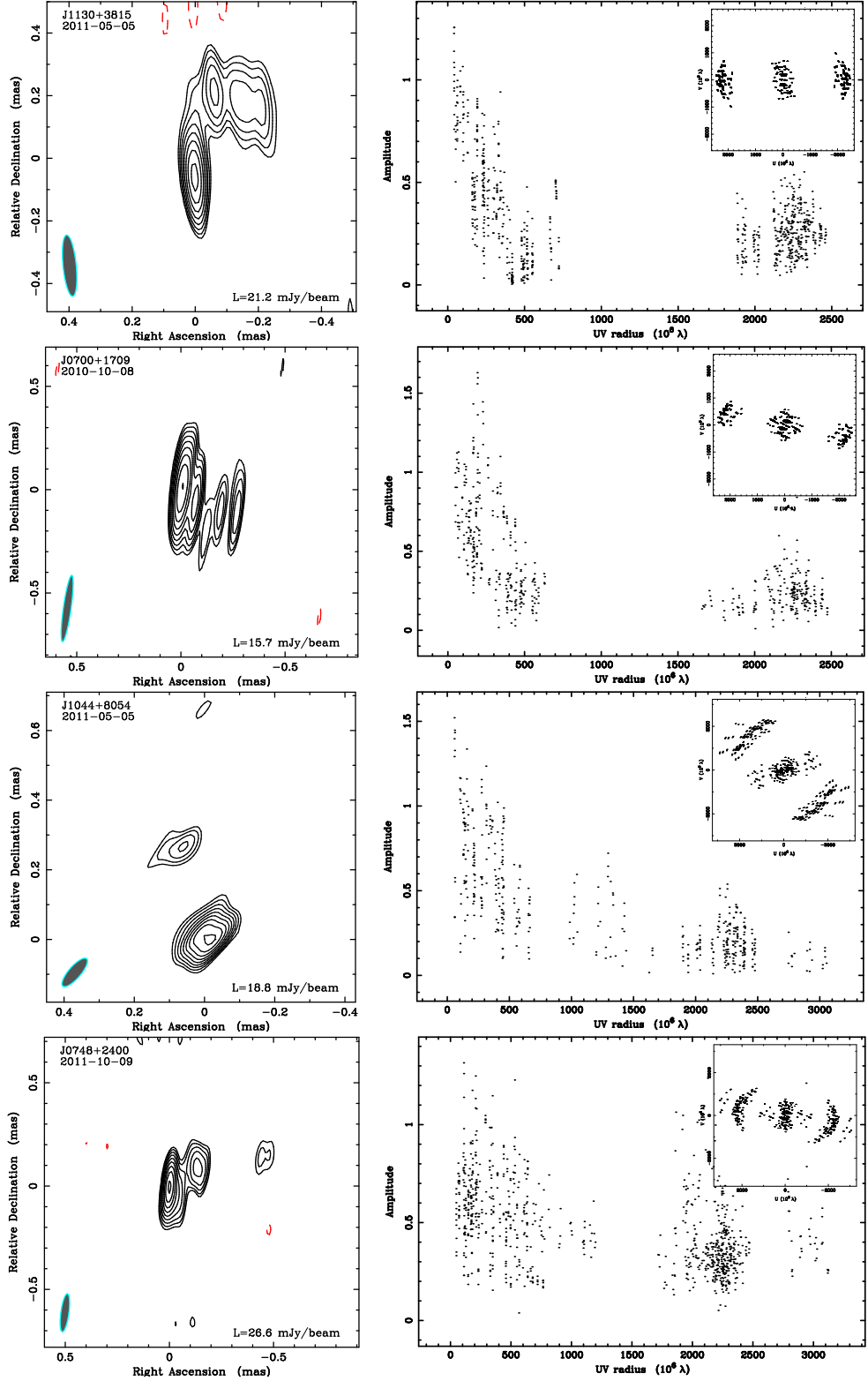

\centerline{\includegraphics[width=0.28\textwidth,angle=-90,trim={1.2cm 0 2.8cm 0},clip=true]{Maps/80_3mm_map_J1130+3815_B.eps}
            \def\big{\includegraphics[width=0.28\textwidth,angle=-90,trim={0.6cm 0 0 0},clip=true]{Maps/80_3mm_radpl_J1130+3815_B.eps}}
            \def\little{\includegraphics[width=0.13\textwidth,angle=-90,trim={0.6cm 0 0 0},clip=true]{Maps/80_3mm_uvpl_J1130+3815_B.eps}}
            \def\stackalignment{r}
            \topinset{\little}{\big}{4pt}{4pt}
}
\centerline{\includegraphics[width=0.28\textwidth,angle=-90,trim={1.2cm 0 2.8cm 0},clip=true]{Maps/43_3mm_map_J0700+1709_A.eps}
            \def\big{\includegraphics[width=0.28\textwidth,angle=-90,trim={0.6cm 0 0 0},clip=true]{Maps/43_3mm_radpl_J0700+1709_A.eps}}
            \def\little{\includegraphics[width=0.13\textwidth,angle=-90,trim={0.6cm 0 0 0},clip=true]{Maps/43_3mm_uvpl_J0700+1709_A.eps}}
            \def\stackalignment{r}
            \topinset{\little}{\big}{4pt}{4pt}
}
\centerline{\includegraphics[width=0.28\textwidth,angle=-90,trim={1.2cm 0 2.8cm 0},clip=true]{Maps/76_3mm_map_J1044+8054_B.eps}
            \def\big{\includegraphics[width=0.28\textwidth,angle=-90,trim={0.6cm 0 0 0},clip=true]{Maps/76_3mm_radpl_J1044+8054_B.eps}}
            \def\little{\includegraphics[width=0.13\textwidth,angle=-90,trim={0.6cm 0 0 0},clip=true]{Maps/76_3mm_uvpl_J1044+8054_B.eps}} 
            \def\stackalignment{r}
            \topinset{\little}{\big}{4pt}{4pt}
}
\centerline{\includegraphics[width=0.28\textwidth,angle=-90,trim={1.2cm 0 2.8cm 0},clip=true]{Maps/51_3mm_map_J0748+2400_C.eps}
            \def\big{\includegraphics[width=0.28\textwidth,angle=-90,trim={0.6cm 0 0 0},clip=true]{Maps/51_3mm_radpl_J0748+2400_C.eps}}
            \def\little{\includegraphics[width=0.13\textwidth,angle=-90,trim={0.6cm 0 0 0},clip=true]{Maps/51_3mm_uvpl_J0748+2400_C.eps}}
            \def\stackalignment{r}
            \topinset{\little}{\big}{4pt}{4pt}
}
\caption{GMVA maps of J1130+3815, J0700+1709, J1044+8054, and J0748+2400 (left panel), shown together with the respective radial
  amplitude distributions (right panel) and {\em uv}-coverages (inset in the right panel) of the respective visibility datasets. The contouring of
  images is made at $3\sigma_\mathrm{rms}\times(-1,1,\sqrt{2},2,...)$ levels, with $\sigma_\mathrm{rms}$ representing the off-source r.m.s noise in the residual image.
  The lowest contour in the maps, L = 21.2~mJy/beam, 15.8~mJy/beam, 18.8~mJy/beam, and 26.6~mJy/beam, respectively. A total of 174 contour maps of 162
  unique sources at 3 mm in this survey are available in the online journal.} 
\label{fig:3mmmaps_main}
\end{figure*}

Using the procedures described above, we have made hybrid maps of
all 174 observations of 162 unique sources in this survey. 
For 138 objects, the survey provides the first ever VLBI images made at 86 GHz.
Most of the imaged sources show extended radio emission, revealing the
jet morphology down to sub-parsec scales.  For a small number of
weaker sources with poor {\em uv}-coverages, only the brightest core
at the base of the jet could be imaged.  To illustrate our results, we
present images of four weak target radio sources 
J1130+3815, J0700+1709, J1044+8054 and J0748+2400 in Figure~\ref{fig:3mmmaps_main}
(the complete figure featuring the full set of images obtained from
this survey is presented in Appendix available electronically).

\begin{figure*}[ht!]
\centerline{\includegraphics[width=0.55\textwidth,trim={0 0 2.35cm 0},clip=true]{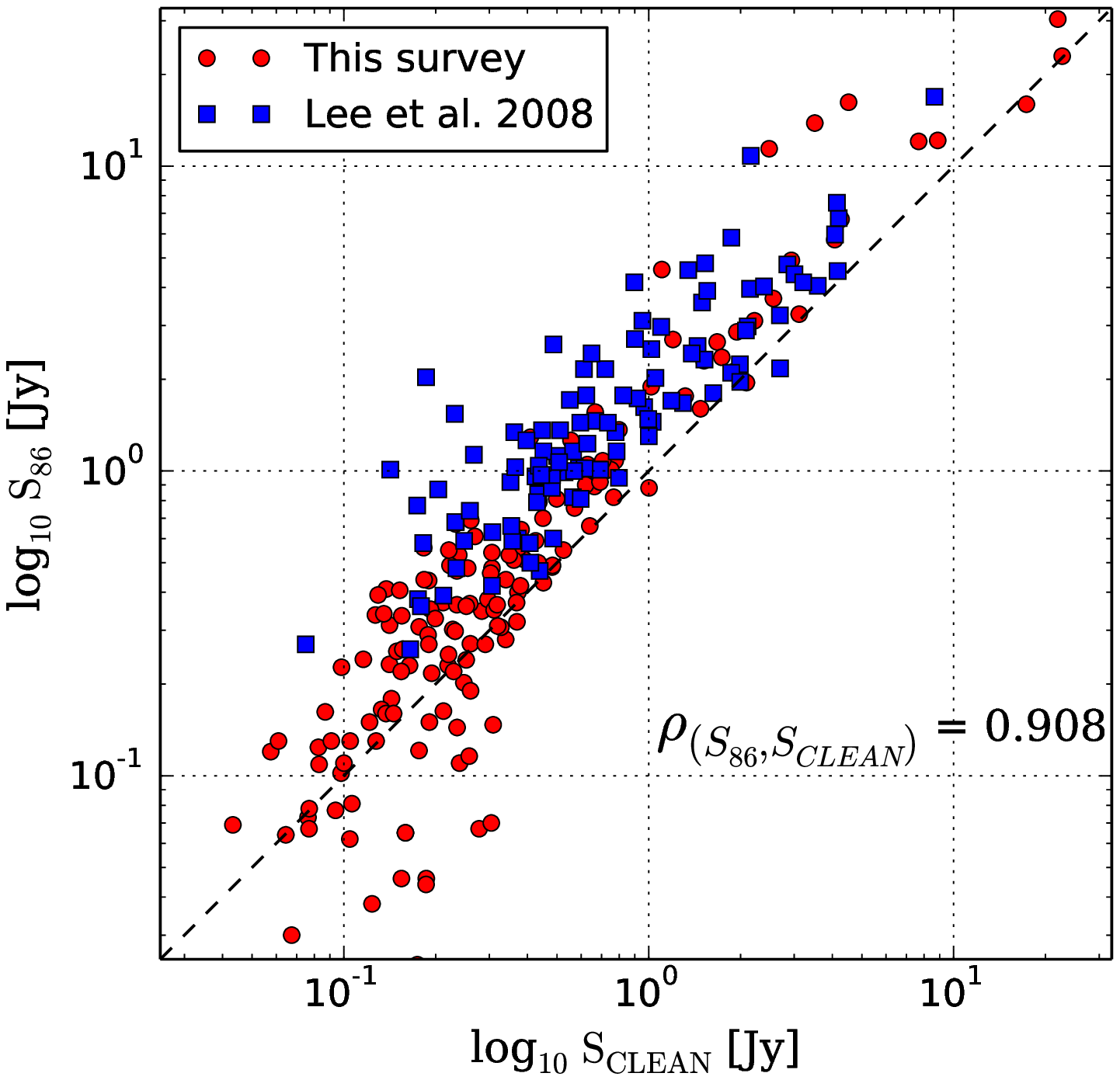}
\includegraphics[width=0.55\textwidth,trim={2.35cm 0 0 0},clip=true]{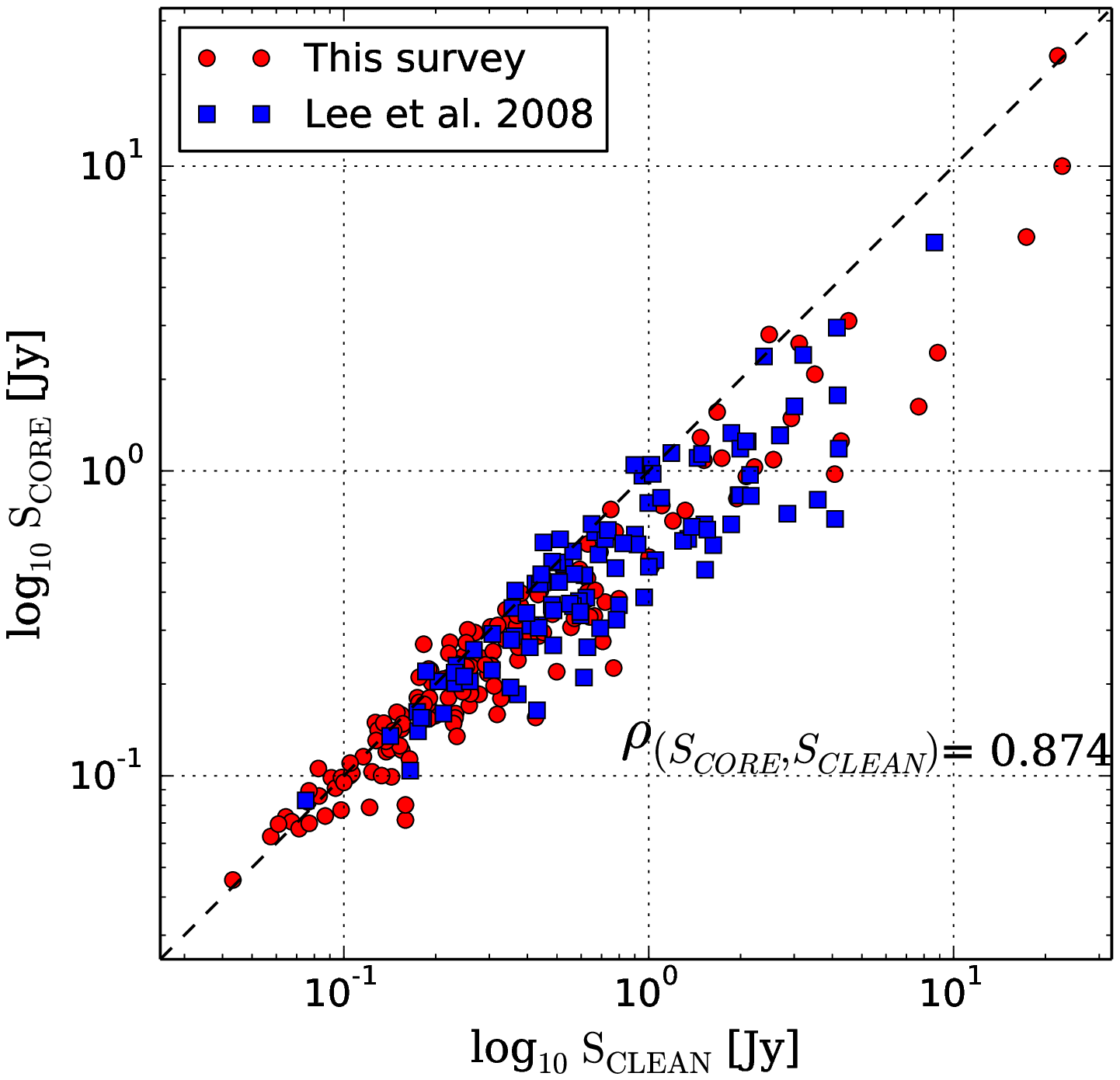}
}
\caption{Compactness parameters, {\normalsize $S_\mathrm{86}$/$S_\mathrm{CLEAN}$} and {\normalsize $S_\mathrm{core}/S_\mathrm{CLEAN}$} 
are shown on the left and right panel respectively, where $S_\mathrm{86}$ is the 
single dish 86\,GHz flux density measured at Pico Veleta or Plateau de Bure. The Pearson 
correlation coefficients { \large $\rho_{({S_\mathrm{86}},{S_\mathrm{CLEAN}})}$} = 0.908 and
{ \large $\rho_{({S_\mathrm{CORE}},{S_\mathrm{CLEAN}})}$} = 0.874 are obtained for the combined
data from this survey and \cite{lee2008}.
}
\label{fig:Scompact}
\end{figure*}

Figure~\ref{fig:Sclean_SL} illustrates the properties of the survey sample by plotting the distributions of the correlated flux density, $S_\mathrm{L}$ ,
measured on long baselines and the total flux density, $S_\mathrm{CLEAN}$,  in the CLEAN images of the observed sources. Both distributions
indicate that objects with flux densities $\gtrsim 80$\,mJy can be successfully detected and imaged with the survey data, signifying the 
sensitivity improvement by a factor of approximately two compared to the observations presented in \cite{lee2008}. 
The mean of the correlated flux density at the longest baseline, $S_\mathrm{L}$ , is
0.2\,Jy. Amongst 157 sources whose $S_\mathrm{L}$ can be measured at projected
baselines longer than 2000 M$\lambda$, 135 sources have an $S_\mathrm{L}$
greater than 0.1 Jy.

In Table~\ref{Image Parameters}, we present the basic parameters of
the images, listing (1)~source name, (2)~observing epoch, (3)~single
dish 86\,GHz flux density, $S_\mathrm{86}$, measured at Pico Veleta or
Plateau de Bure, (4)~correlated flux density on the shortest baseline,
$S_\mathrm{S}$, (5)~shortest baseline, $B_\mathrm{S}$, (6)~correlated
flux density on the longest baseline, $S_\mathrm{L}$, (7)~longest
baseline, $B_\mathrm{L}$, (8)~major axis, $B_\mathrm{a}$, (9)~minor
axis, $B_\mathrm{b}$, and (10)~position angle, $B_\mathrm{PA}$ of the
major axis of the restoring beam, (11)~total CLEAN flux density,
$S_\mathrm{tot}$, and (12)~peak flux density, $S_\mathrm{peak}$, in the
image, (13)~image r.m.s noise, $\sigma_\mathrm{rms}$, and (14)~the
quality factor of the residual noise in the image, $\xi_\mathrm{r}$ .

Table~\ref{Model fit parameters} summarizes the model fits obtained
for all of the survey sources providing (1)~the source name,
(2)~observing epoch, (3)~sequential number of the Gaussian component,
(4)~total flux density, $S_\mathrm{t}$, and (5)~peak flux density,
$S_\mathrm{peak}$, of the component, (6)~size, $d$, of the component,
(7)~separation, $r$, and (8)~position angle, $\theta$ of the component
with respect to the brightest feature in the model (core, taken to
be located at the coordinate origin), (9)~brightness temperature,
$T_\mathrm{b,mod}$, estimated from the model fit, and (10)~minimum,
$T_\mathrm{b,min}$ and (11)~limiting resolved, $T_\mathrm{b,lim}$,
brightness temperatures estimated from the visibility amplitudes
\citep{lobanov2015} at the longest baselines given in Column 7 in 
Table~\ref{Image Parameters}.

Table~\ref{Model fit parameters} contains the model fit parameters for a total 174 VLBI cores and 205
jet components, with 42 and 37 of these unresolved, as reflected also
in the lower limits of the model-fit-based brightness temperature
estimates, $T_\mathrm{b,mod}$, listed for these components.

\subsection{Source compactness} 

Compactness of the source structure can be evaluated by comparing the
single dish flux density, $S_\mathrm{86}$ , listed in Table~\ref{Image Parameters}, to the total clean flux density, $S_\mathrm{CLEAN}$ , 
listed in Table~\ref{Image Parameters} and the core flux density, $S_\mathrm{CORE}$ , listed in Table~\ref{Model fit parameters},
to the total clean flux density, $S_\mathrm{CLEAN}$. These comparisons are presented in Figure~\ref{fig:Scompact}.

To study the relation between the total and VLBI flux densities, we apply the Pearson correlation test.
The Pearson correlation coefficient calculated for $S_\mathrm{86}$ and $S_\mathrm{CLEAN}$ gives a significant value of 0.924 for this survey and 
0.908 for this survey combined with the results from \cite{lee2008}. The respective plot in the left panel of Figure~\ref{fig:Scompact}
indicates that almost all the flux measured by a single dish (here Pv or PdB)
is recovered in the VLBI clean flux. The median of the core dominance index defined as $S_\mathrm{CORE}$/$S_\mathrm{CLEAN}$
is 0.84, and the two are also correlated, as demonstrated in the right panel of Figure~\ref{fig:Scompact}.

Figure~\ref{fig:Scompact} shows that the stronger sources have more structures (right panel) some of which are completely resolved out even on the shortest baselines of the 
survey observations (left panel). A small number of cases for which $S_\mathrm{CLEAN}$ > $S_\mathrm{86}$ or $S_\mathrm{CORE}$ > $S_\mathrm{CLEAN}$ are observed for
the weaker objects. These can be reconciled with the errors in the measurements, and they essentially imply very compact objects, 
with $S_\mathrm{86}$ $\simeq$ $S_\mathrm{CLEAN}$ and $S_\mathrm{CORE}$ $\simeq$ $S_\mathrm{CLEAN}$, respectively.

\subsection{Brightness temperatures}
\label{sc:tb}

In our further analysis, we use the model-fit-based and
visibility-based estimates of the brightness temperature of the VLBI
bright core (base) and the inner ($r_\mathrm{proj}<1$\,pc) jet
components, taking into account the resolution limits of the data at
3\,mm. The brightness temperature estimates for the jet cores in our sample 
range from $2.5 \times 10^{9}$\,K to $ 1.3\times 10^{12}$\,K.
The brightness temperature estimates for the inner jet components in our sample 	
range from $7.0 \times 10^{7}$\,K to $4.0 \times 10^{11}$\,K.  
The median and mean of the brightness temperature distribution
for the core regions are $8.6 \times 10^{10}$\,K and $1.8 \times
10^{11} $\,K, respectively. For the inner jet components, the
respective figures are $7.2 \times 10^{9}$\,K and $2.2 \times
10^{10}$\,K. This shows that the brightness temperature drops by
approximately an order of magnitude already on sub-parsec scales in
the jets, with inverse Compton, synchrotron, and adiabatic losses subsequently
dominating the energy losses \citep[cf.][]{marscher1995,lobanov1999}.
Only 8\,$\%$ of the jet cores show a brightness temperature greater than
$5 \times 10^{11}$\,K and only 3\,$\%$ have a brightness
temperature greater than $10^{12}$\,K.

We also inspect the distribution of the minimum and maximum limiting
brightness temperature of the core components (using averaged values
of brightness temperature for objects with multiple observations) in
the sample, making these estimates from the visibility amplitudes on
the longest baselines \citep{lobanov2015}. The minimum, $T_\mathrm{b,min}$ ,
and limiting, $T_\mathrm{b,lim}$ , brightness temperatures are 
given in Table 7, in columns 10 and 11, respectively.

\begin{figure}[ht!]
  \centerline{\includegraphics[width=0.5\textwidth,trim={0 0 0 1.35cm},clip=true]{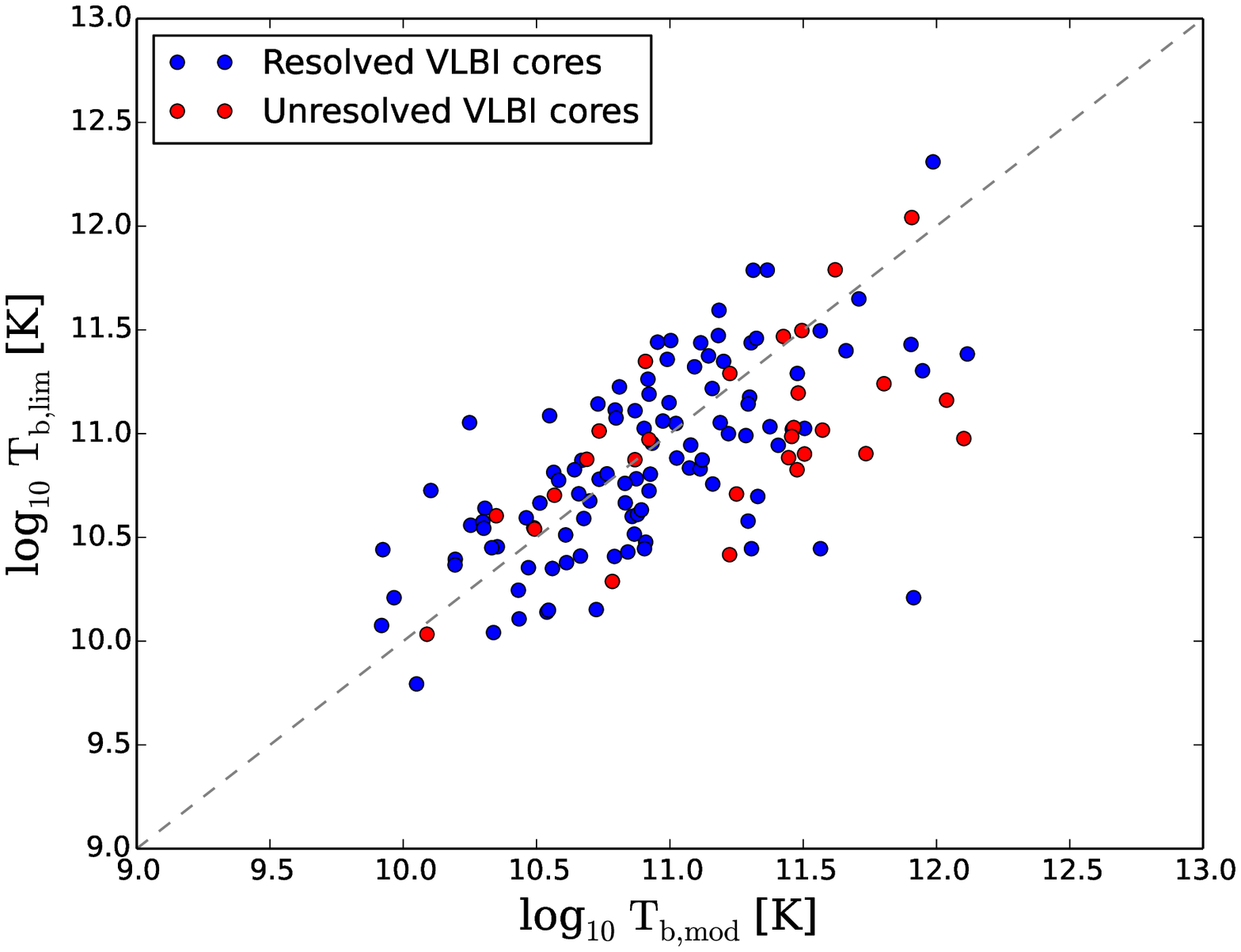}}
\centerline{\includegraphics[width=0.5\textwidth]{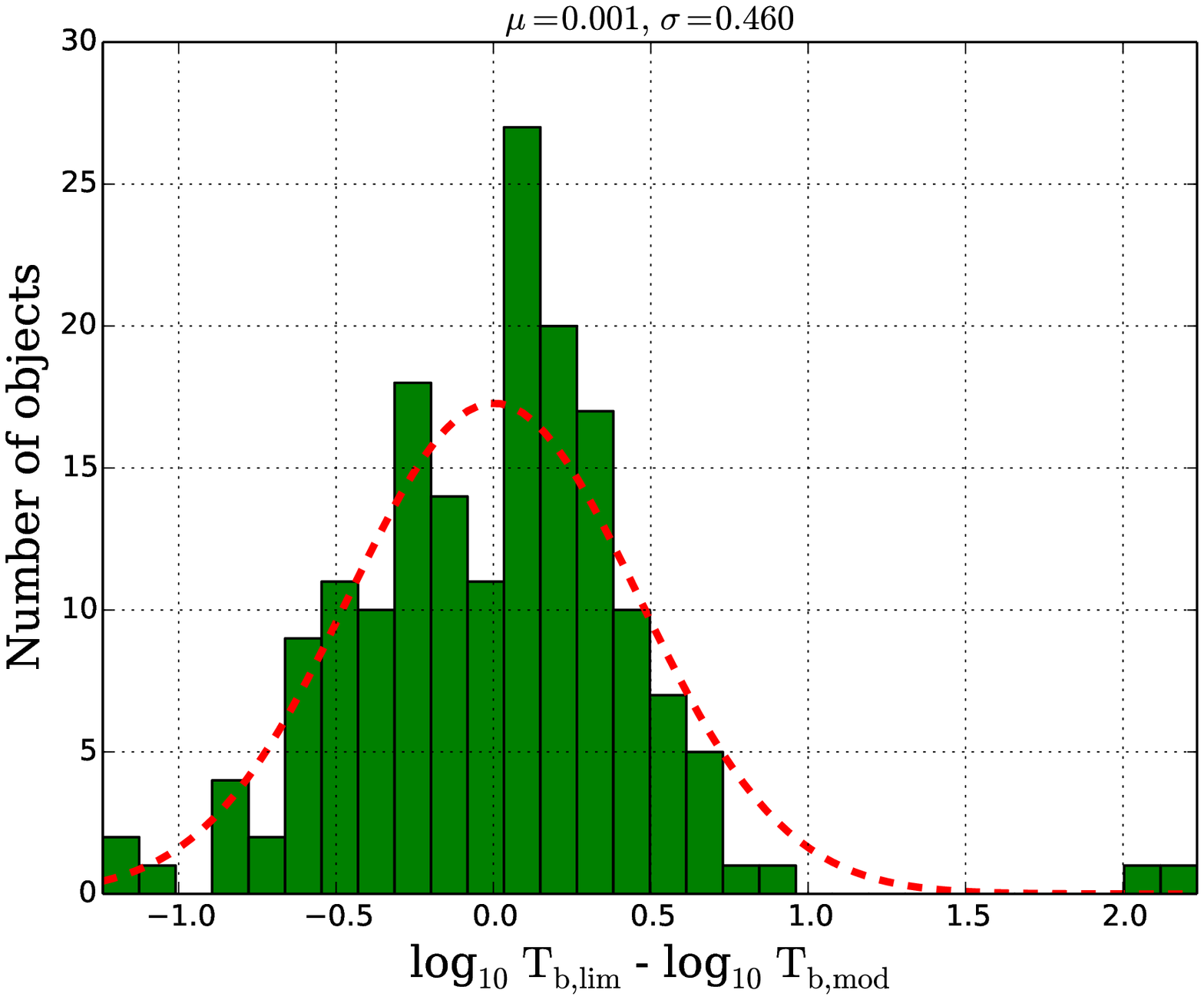}
\label{Tbmod_Tblim}              
}
\caption{Comparison of $T_\mathrm{b,mod}$ measured from circular
  Gaussian representation of source structure and $T_\mathrm{b,lim}$
  estimated from the interferometric visibilities at {\em uv}-radii
  within 10\% of the maximum baseline $B_\mathrm{max}$ in the data for
  a given source (top panel). Correlation between the two distributions
  is illustrated by the residual logarithmic distribution of the
  $T_\mathrm{b,mod}$/$T_\mathrm{b,limit}$ ratio (bottom panel), which
  is well approximated by the Gaussian distribution with $\mu$ = 0.001
  and $\sigma$ = 0.46.}
\label{fg:tbvisibility}
\end{figure}

The median and mean of the maximum
limiting brightness temperature distribution
for the core regions is $1.06 \times 10^{11}$\,K and $3.0 \times
10^{11} $\,K, respectively. We find that the limiting
$T_\mathrm{b,lim}$ correlates well with $T_\mathrm{b,mod}$ estimated
from imaging method as seen in Figure~\ref{fg:tbvisibility},
supporting the fidelity of $T_\mathrm{b,mod}$ measurements obtained
from model fitting. The residual logarithmic distribution of the
$T_\mathrm{b,mod}$/$T_\mathrm{b,lim}$ ratio is well approximated by
the Gaussian PDF, with mean value, $\mu$ = 0.001 and standard deviation, 
$\sigma$ = 0.46.

\section{Discussion}
\label{sc:discussion}

\subsection{Modelling the observed brightness temperatures}
\label{sc:model}

The brightness temperature distribution can be used for obtaining
estimates of the conditions in the extra galactic radio sources and to
test the models proposed for the inner jets
\citep{marscher1995,lobanov2000,homan2006,lee2008}. A basic population
model \citep{lobanov2000} can be used for representing the observed
brightness temperature distribution under the assumption that the jets
have the same intrinsic brightness temperature, $T_\mathrm{0}$, Lorentz factor, $\Gamma_\mathrm{j}$, and synchrotron spectral index, $\alpha$
($S_\nu \propto \nu^\alpha$), and they are 
randomly oriented in space (within the limits of viewing angles, $\theta$ , required by Doppler boosting bias).
The jets are also assumed to remain straight within the spatial scales ($\sim 0.5-10\,\mathrm{pc}$) probed by the observations.

The assumptions of single values of $T_\mathrm{0}$ and $\Gamma_\mathrm{j}$ describing the whole sample are clearly simplified, as jets are known to 
feature a range of Lorentz factors \citep[see][and references therein]{lister2016}. However, as has been shown earlier \citep{lobanov2000},
factoring a distribution of Lorentz factors into the present model is not viable without amending the brightness temperature measurements with additional 
information, preferably about the apparent speeds of the target sources. We are currently compiling such a combined database, and will engage in a more 
detailed modelling of the compact jets after the completion of this database.

In a population of jets described by the settings summarized above, the measured
brightness temperature, $T_\mathrm{b}$, is determined solely by the
relativistic Doppler boosting of the jet emission.
Therefore, the observed brightness temperature , $T_\mathrm{b}$ , can be related to the intrinsic brightness temperature, 
$T_0$, so that $T_\mathrm{b} = T_0\,\delta^{1/\epsilon}$, where the power index $\epsilon$ is $1/(2-\alpha)$ for  a continuous jet (steady state jet)
and $1/(3-\alpha)$ for a jet with spherical blobs (or optically thin ``plasmoids''), 
and $\delta$ is the Doppler factor.

The probability of finding a radio source with the brightness temperature
$T_\mathrm{b}$ in such a population of sources is
\begin{equation}
\label{eqn:prob_equation}
p(T_\mathrm{b}) \propto \Bigg[ \frac{2\,\Gamma_{j}\left(T_\mathrm{b}/{T_\mathrm{0}}\right)^\epsilon - \left(T_\mathrm{b}/{T_\mathrm{0}}\right)^{2\epsilon} - 1} {\Gamma_\mathrm{j}^{2} - 1}\Bigg]^{\frac{1}{2}}\,.
\end{equation}

\begin{figure}[ht!]
\centering
\includegraphics[width=0.52\textwidth,trim={1cm 0 1cm 1.3cm},clip=true]{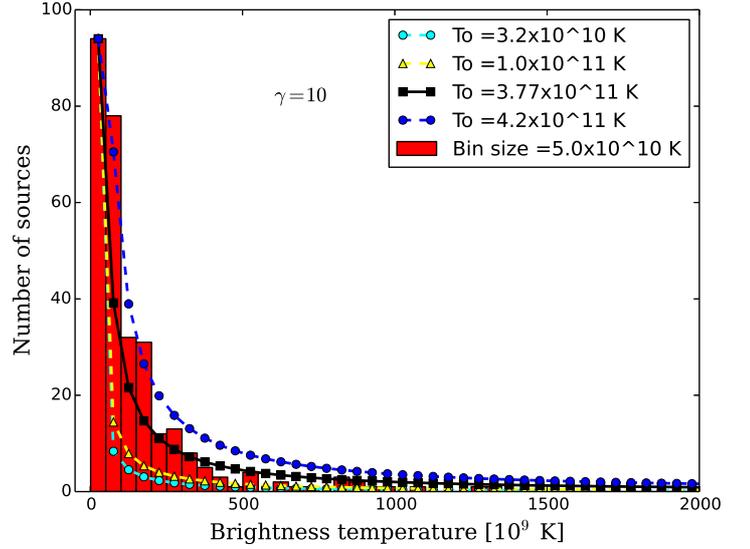}
\caption{Distribution of the brightness temperatures, $T_\mathrm{b}$ , measured in the core components and represented by the population models calculated for 
$\Gamma_\mathrm{j}=10$ and different values of $T_\mathrm{0}$. 
The best approximation of the observed $T_\mathrm{b}$ distribution is obtained
with $T_\mathrm{o,core}$ = $(3.77^{+0.10}_{-0.14}) \times 10^{11}$\,K. For better viewing of the observed distribution, one core component with a 
very high $T_\mathrm{b}$ = $ 5.5 \times 10^{12}$\,K for the source BL Lac obtained from \cite{lee2008}
is not shown but is included in the modelling.}
\label{fg:tbcores_all}
\end{figure}

\begin{figure}[ht!]
\centering
\includegraphics[width=0.52\textwidth,trim={1cm 0 1cm 1.3cm},clip=true]{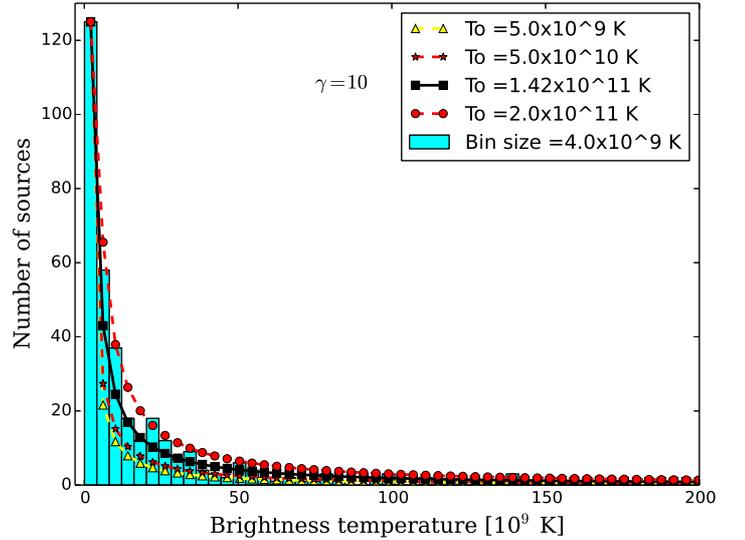}
\caption{Distribution of the brightness temperatures, $T_\mathrm{b}$ ,
  measured in the inner jet components and represented by the population
  models calculated for $\Gamma_\mathrm{j}=10$ and different values of
  $T_\mathrm{0}$. The best approximation of the observed $T_\mathrm{b}$ distribution is obtained
  with $T_\mathrm{o,jet} = (1.42^{+0.16}_{-0.19}) \times 10^{11}$\,K.
  }
\label{fg:tbjetcomponents_all}
\end{figure}

The lower end of the observed distribution of brightness temperatures depends on the sensitivity of VLBI data since the flux of the
observed sample is biased by Doppler boosting \citep{lobanov2000}.
The lowest brightness temperature that can be measured from our data,
$T_\mathrm{b,sens}$ , can be obtained from
\begin{equation}
\label{eqn:Tb_sens}
T_{\mathrm{b,sens}} \mathrm{[K]} = 1.65 \times 10^{5} \left(\frac{\sigma_{\mathrm{rms}}}{\mathrm{mJy/beam}}\right)  \left(\frac{b}{\mathrm{mas}}\right)^{-2}\,,
\end{equation}

\noindent where $\sigma_\mathrm{rms}$ is the array sensitivity in mJy/beam and $b$ is the average size of the resolving beam.
In this survey, the typical observation time on a target source $\Delta t$ is 20 minutes 
and bandwidth is 128\,MHz.
Therefore, the value of beam size for the sources in this survey is 0.12 mas
and the $\sigma_\mathrm{rms}$ of the array is 0.54 mJy/beam.
Thus, we have obtained a 3$\sigma$ level estimation of $T_\mathrm{b,sens}$ as 2.0 $\times$ 10$^{8}$ K using 
Equation~\ref{eqn:Tb_sens}, which is set as the lowest brightness temperature in modelling.

We normalize the results obtained from Equation (\ref{eqn:prob_equation}) to the
number of objects in the lowest bin of the histogram.
For our modelling, we first make a generic assumption of $\alpha=-0.7$
(homogeneous synchrotron source) and use $\Gamma_\mathrm{j} \approx 10$
implied from the kinematic analysis of the MOJAVE VLBI survey
data \citep{lister2016}. We assume that the jet is continuous, so $\epsilon =$ 0.37 is taken. 
For the population modelling analysis, we have included the data 
from \cite{lobanov2000}, \cite{lee2008},
and the present survey, yielding a final database of 271
VLBI core components and 344 jet components. For objects with multiple measurements, we 
have used the median of the measurements.
The resulting model distributions obtained for
various values of $T_\mathrm{0}$ are shown in
Figures~(\ref{fg:tbcores_all}--\ref{fg:tbjetcomponents_all}) for the VLBI
cores and the inner jet components, respectively.

This approach yields $T_\mathrm{0,core}$ = $(3.77^{+0.10}_{-0.14}) \times 10^{11}$\,K
for the VLBI cores and $T_\mathrm{0,jet} = (1.42^{+0.16}_{-0.19})
\times 10^{11}$\,K for the inner jet components. 
The estimated $T_\mathrm{0,core}$ is in good agreement with the inverse Compton
limit of $\simeq$ 5.0 $\times$ 10$^{11}$\,K \citep{kellermann1969},
beyond which the inverse Compton effect causes rapid electron energy
losses and extinguishes the synchrotron radiation. 
The inferred $T_\mathrm{0,jet}$ of jet components are 
about a factor of three higher than the
equipartition limit of $\simeq$ 5 $\times$ 10$^{10}$\,K
\citep{readhead1983} for which the magnetic field energy and particle
energy are in equilibrium. This may indicate that opacity is still non-negligible in these regions of the flow.
The intrinsic brightness temperature obtained for the cores is within the upper limit $5.0 \times 10^{11}$\,K predicted 
for the population modelling of the cores \citep{lobanov2000}.

\begin{figure*}[ht!]
\centering
\includegraphics[width=0.90\textwidth,clip=true]{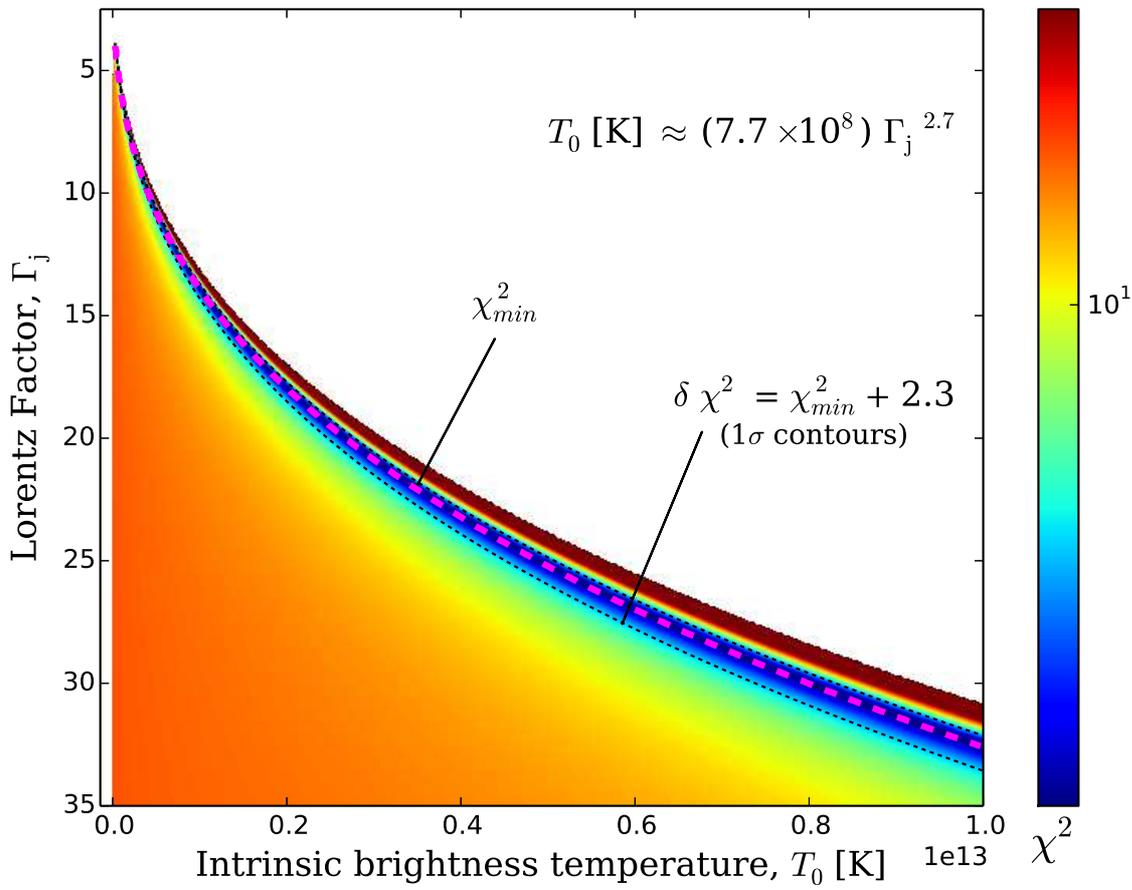}
\caption{Two-dimensional
  ${\chi}^2$ distribution plot in the $\Gamma_\mathrm{j}$ -- $T_\mathrm{0}$
  space, calculated for the brightness temperatures measured in the VLBI cores.
  The blank area shows the ranges of the parameter space
  disallowed by the observed distribution.  The distribution of the
  ${\chi}^2$ values indicates a ($\Gamma_\mathrm{j}$--$T_\mathrm{0}$) correlation,
  with $T_\mathrm{0}[\mathrm{K}]\approx (7.7\times
  10^{8})\,\Gamma_\mathrm{j}^{2.7}$, thus precluding a simultaneous
  fit for $\Gamma_\mathrm{j}$ and $T_\mathrm{0}$.}
\label{fg:simfit}
\end{figure*}

A simultaneous fit for $T_\mathrm{0}$ and $\Gamma_\mathrm{j}$ is
impeded by the implicit correlation, $T_\mathrm{0} \propto \Gamma_\mathrm{j}^{a}$
(with $a \approx 2$--3), between these two parameters, as implied by
Equation (\ref{eqn:prob_equation}). This is also illustrated in
Figure~\ref{fg:simfit}, from which a dependence $T_\mathrm{0}[\mathrm{K}] \approx
(7.7\times 10^{8})\,\Gamma_\mathrm{j}^{2.7}$ can be
inferred for the fit to the brightness temperatures measured in the VLBI cores.
This correlation between $T_\mathrm{0}$ and $\Gamma_\mathrm{j}$ precludes 
simultaneously fitting for both these parameters, and hence the 

\noindent Lorentz factor has to be constrained (or assumed) separately. One should also keep in mind that this correlation results from the model
description and does not have an immediate physical implication.  
Equation (\ref{eqn:prob_equation}) clearly shows that the predicted distribution of $T_\mathrm{b}$ is valid within the range

\begin{equation}
\label{eqn:prob_range}
(\Gamma_\mathrm{j} - \sqrt{ \Gamma_\mathrm{j}^{2} - 1}) \leq (\frac{T_\mathrm{0}}{T_\mathrm{b}})^{\epsilon}  \leq (\Gamma_\mathrm{j} + \sqrt{ \Gamma_\mathrm{j}^{2} - 1})\,.
\end{equation}
The region outside this range is represented by the blank area in Figure~\ref{fg:simfit}. 

The intrinsic brightness temperature we obtained is higher than the mean and median observed brightness temperature $T_\mathrm{b}$.
This is readily explained by the Doppler deboosting. 
For a given viewing angle, $\theta$, sources with $\Gamma_\mathrm{j}>1/\theta$ would be deboosted so that the observed 
brightness temperature will be reduced below its intrinsic value. It can be easily shown that the observed and
intrinsic brightness temperatures are equal if the jet viewing angle is given by
\begin{equation}
\label{eqn:viewing angle}
\theta_\mathrm{eq} = \arccos \left[ \frac{1- (1/\Gamma_\mathrm{j})\,\,(T_\mathrm{0}/T_\mathrm{b})^{\epsilon}} {\sqrt{1-\Gamma_\mathrm{j}^{-2}}} \right]\,.
\end{equation}
For the VLBI cores, the mean of the observed $T_\mathrm{b}$ is $1.8 \times 10^{11}$\,K and intrinsic $T_\mathrm{0,core}$ is $3.77 \times 10^{11}$\,K, 
therefore, the resulting $\theta_\mathrm{eq}$ = 29$^{\circ}$ for $\Gamma_\mathrm{j}$ = 10 and $\epsilon$ = 0.37.
In this case any object observed at a larger viewing angle would be deboosted resulting in a lower observed $T_\mathrm{b}$ than intrinsic $T_\mathrm{0}$.

\subsection{Testing the adiabatic expansion of jets}
\label{sc:evol}

As discussed in Sections~\ref{sc:tb} and \ref{sc:model}, 
intrinsic $T_\mathrm{0}$ and the observed $T_\mathrm{b}$ in core and jets show that the brightness temperature drops
by approximately a factor of two to ten already on sub-parsec scales in
the jets. This evolution might occur with the inverse Compton, synchrotron, and adiabatic losses subsequently
\begin{figure*}[ht!]
\centerline{\includegraphics[width=0.5\textwidth]{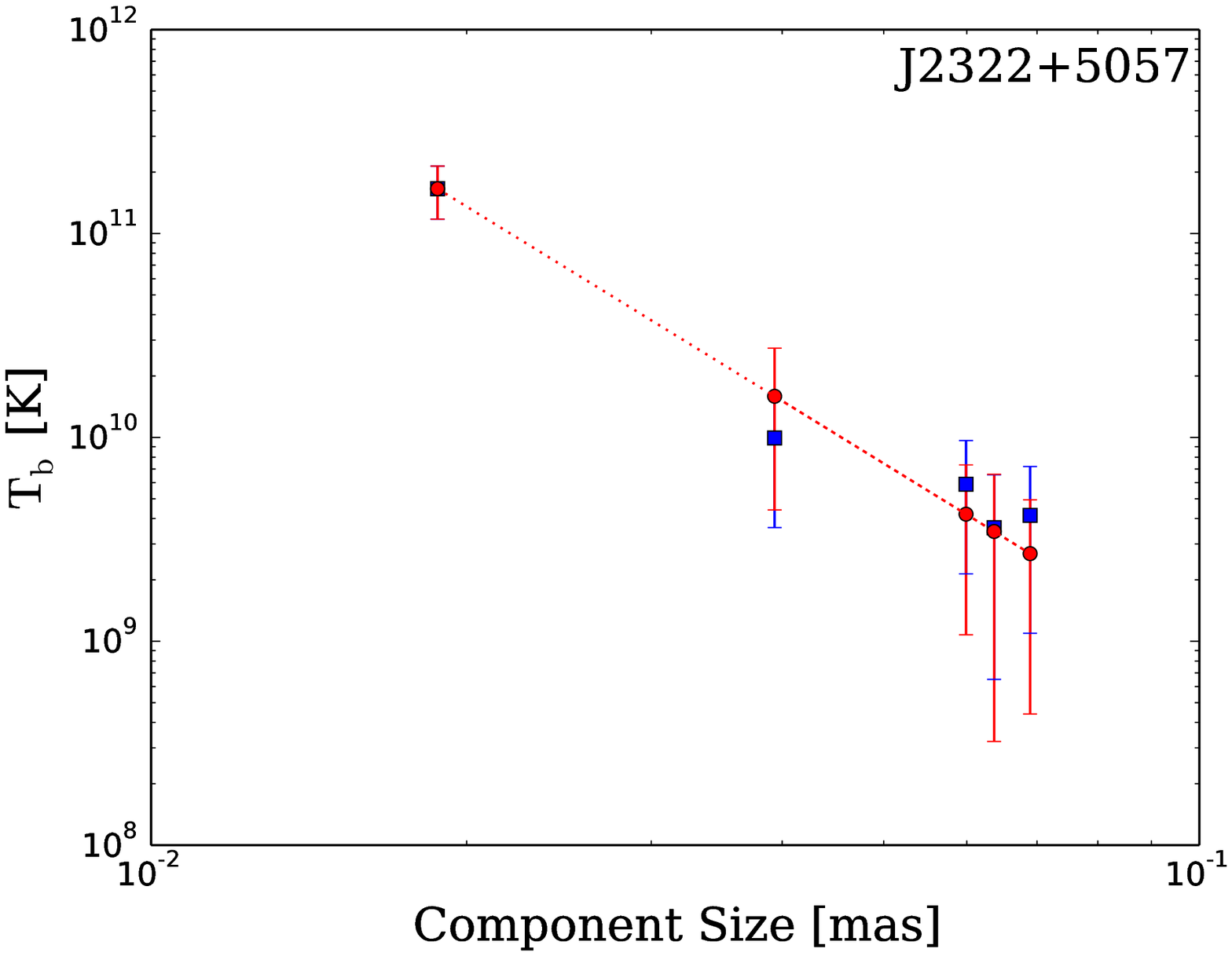}
            \includegraphics[width=0.5\textwidth]{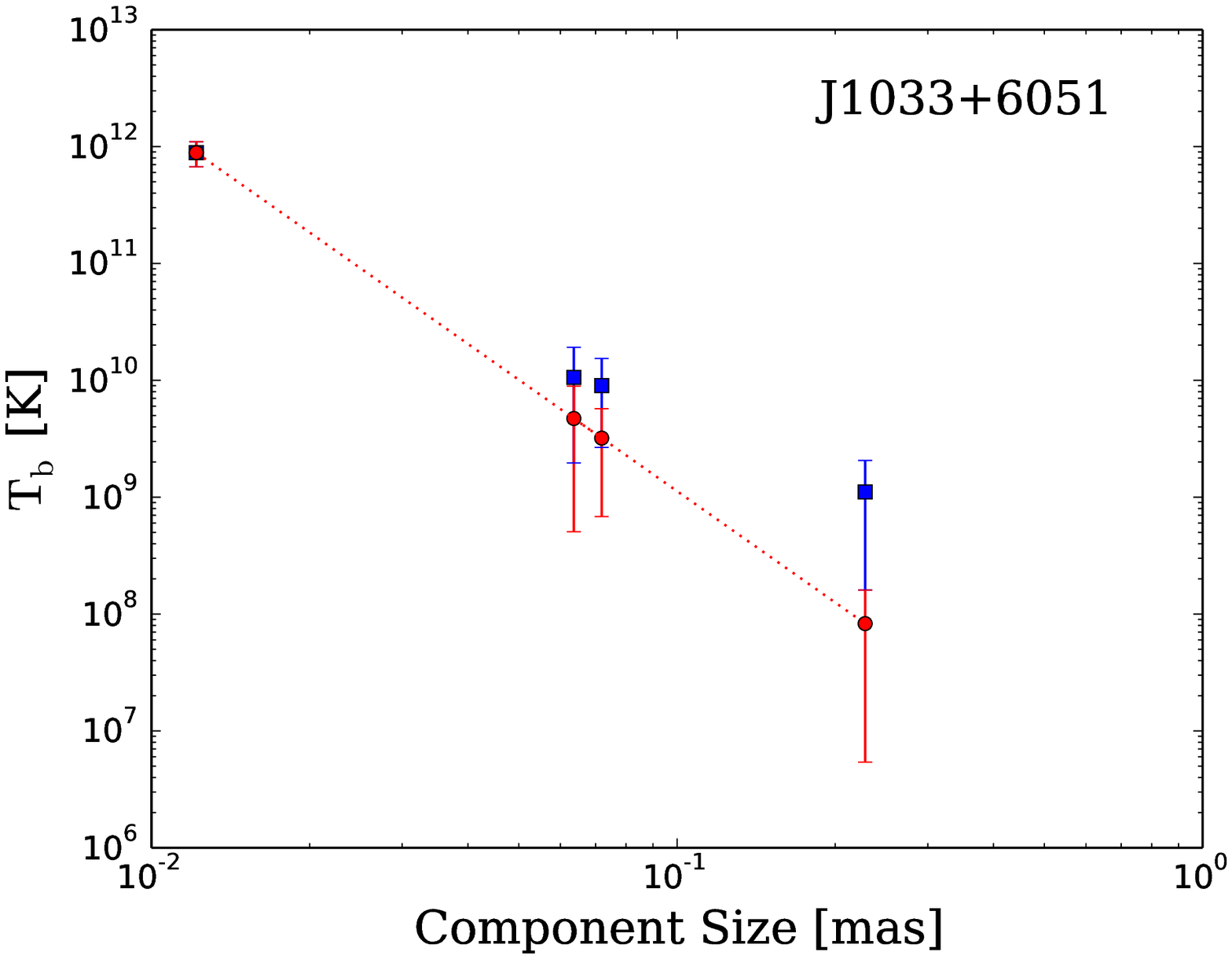}
}
\centerline{\includegraphics[width=0.5\textwidth]{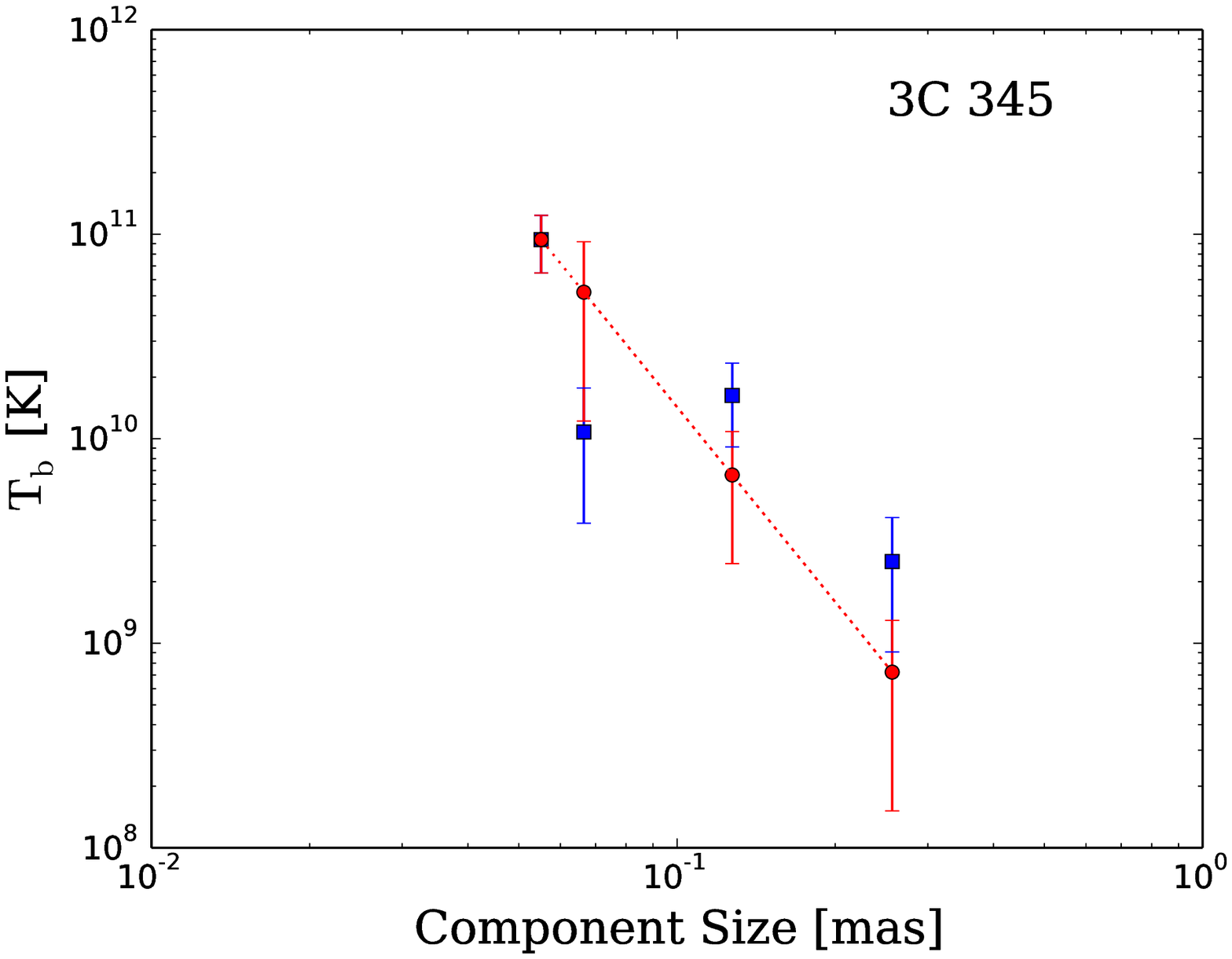}
            \includegraphics[width=0.5\textwidth]{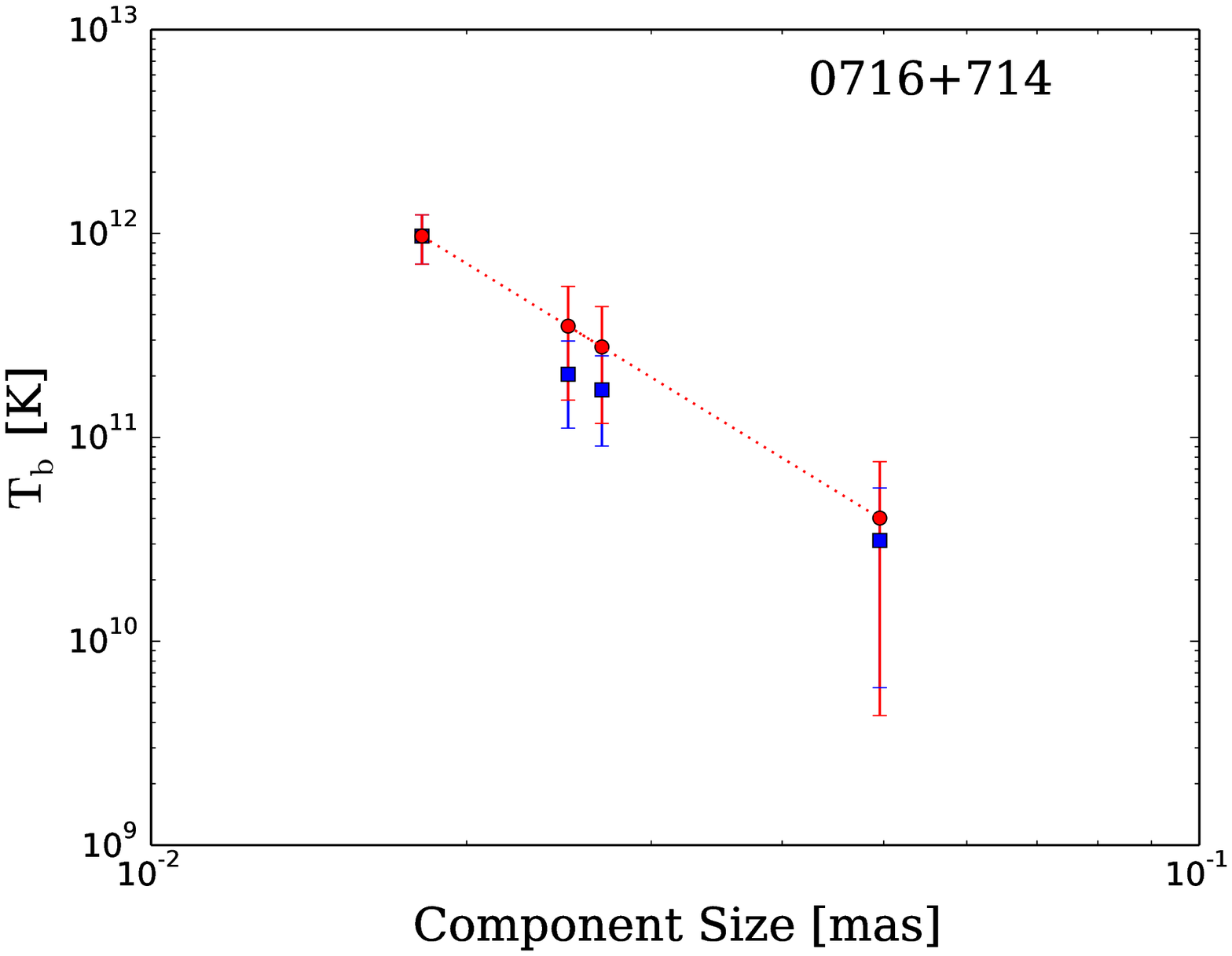}
}
  \caption{Changes of the brightness temperature as a funcion of jet width for four sources --  J2322+5037, J1033+6051, 3C 345, and 0716+714 from this survey. 
Blue squares denote the measured $T_\mathrm{b}$ from this survey. The red circles connected with a dotted line represent
theoretically expected $T_\mathrm{b}$ under the assumption of adiabatic jet expansion. The initial brightness temperature in each jet is 
assumed to be the same as that measured in the VLBI core.}
\label{fig:adiabatic}
\end{figure*}
dominating the energy losses \citep[cf.,][]{marscher1995,lobanov1999}.

For four objects in our data (3C84, 0716+714, 3C454.3, and J2322+507)
for which multiple jet components have been identified during the
model fitting, it is possible to use the brightness temperatures of the
jet components to test whether the evolution of the jet brightness on sub parsec scales
could be explained by adiabatic energy losses
\citep{marscher1985}. For this analysis, we assume that the jet
components are independent relativistic shocks embedded in the jet
plasma, which has a power-law distribution $ N(E)\,\mathrm{dE} \propto
E^{-s}\mathrm{dE,}$ where $s$ is the energy spectral index 
that depends on spectral index $\alpha$ as $\alpha = (1-s)/2 $, 
and is pervaded by the magnetic field $B \propto$
$d^{-a}$, where $d$ is the width of the jet and $a$ depends on the 
type of magnetic field ($a =$ 1 for poloidal magnetic field and 2 for 
toroidal magnetic field). With 
these assumptions, we can relate the brightness temperatures, $T_\mathrm{b,J}$, of the
jet components to the brightness temperature, $T_\mathrm{b,C}$, of the
core \citep{lobanov2000,lee2008},
\begin{equation}
\label{eqn:adiab}
T_\mathrm{b,J} = T_\mathrm{b,C} \left(\frac{d_\mathrm{J}}{d_\mathrm{C}}\right)^{-\xi}\,,
\end{equation}
where $d_\mathrm{J}$ and $d_\mathrm{C}$ are the measured sizes of the jet component and core, respectively, and 
\begin{equation}
\xi = \frac {2(2s+1)+3a(s+1)}{6}\,.
\end{equation}

Assuming the synchrotron emission with spectral index $\alpha$ = -0.5,
we use $s = 2.0$ and adopt $a = 1.0$ for the description of
the magnetic field in the jet. With these assumptions, we calculate
the predicted $T_\mathrm{b,J}$ for individual jet components and
compare them in Figure~\ref{fig:adiabatic} to the measured brightness
temperatures. The measured and predicted values of brightness temperature
agree well, and this suggests that the jet components can be viewed as adiabatically
expanding relativistic shocks \citep[cf.,][]{kadler2004,pushkarev2012,kravchenko2016}.

\section{Summary}

We have used the Global Millimeter VLBI Array (GMVA) to conduct a
large global 86 GHz VLBI survey comprising 174 snapshot observations
of 162 unique targets selected from a sample of compact radio sources. The survey 
observations have reached a typical baseline sensitivity of
0.1\,Jy and a typical image sensitivity of 5\,mJy/beam,
owing to the increased recording bandwidth of the GMVA observations
and the participation of very sensitive European antennae at Pico Veleta and
Plateau de Bure. All of the 162 objects have been detected and
imaged, thereby increasing the total number of
AGN imaged with VLBI at 86$\,$GHz by a factor of $\sim 1.5$. We imaged
138 sources for the first time with VLBI at 86\,GHz through this survey.

We have used Gaussian model fitting to represent the structure of the
observed sources and estimate the flux densities and sizes of the core
and jet components. We used the model fit parameters and visibility data on the longest
baselines to make independent estimates of brightness temperatures at
the jet base as described by the most compact and bright ``VLBI core''
component. These estimates are consistent with each
other. For sources with extended structure detected, the model fit
parameters have been also used to calculate brightness temperature
in the jet components downstream from the core. 
The apparent brightness temperature estimates for the jet cores in our sample 
range from $2.5 \times 10^{9}$\,K to $ 1.3\times 10^{12}$\,K, with the mean value of $1.8 \times 10^{11} $\,K.
The brightness temperature estimates for the inner jet components in our sample 
range from $7.0 \times 10^{7}$\,K to $4.0 \times 10^{11}$\,K.  
The overall amplitude calibration error for the observations is about $25\%$.
    
We describe the observed brightness temperature distributions by a
basic population model which assumes that all jets are intrinsically
similar and can be described by a single value of the intrinsic
brightness temperature, $T_\mathrm{0}$, and Lorentz factor,
$\Gamma_\mathrm{j}$. The population modelling shows that our data are
consistent with a population of sources that has $T_\mathrm{0} = (3.77^{+0.10}_{-0.14}) \times 10^{11}$\,K 
in the VLBI cores and  $T_\mathrm{0} = (1.42^{+0.16}_{-0.19}) \times 10^{11}$\,K in the jets, both obtained for
$\Gamma_\mathrm{j} = 10$ adopted from the kinematic analysis of the
MOJAVE VLBI survey of AGN jets \citep{lister2016}. A correlation
between $T_\mathrm{0}$ and $\Gamma_\mathrm{j}$ inherent to the model
description precludes fitting for these two parameters
simultaneously. We find that a relation $T_\mathrm{0}[\mathrm{K}] \approx (7.7
\times 10^{8})\,\Gamma_\mathrm{j}^{2.7}$ is implied for
this modelling framework by the survey data.
For sources with sufficient structural detail, there is an agreement
between the brightness temperatures measured in multiple components along
the jet and the predicted brightness temperatures for relativistic 
shocks with adiabatic losses dominating the emission. 

The results of the survey can be combined with brightness temperature
measurements made from VLBI observations at lower frequencies \citep[{
e.g.},][]{kovalev2005,petrov2007} to study the evolution of
$T_\mathrm{o}$ with frequency and along the jet
\citep{lee2008,lee2016}. This approach can be used to better constrain
the bulk Lorentz factor and the intrinsic brightness temperature, to
distinguish between the acceleration and deceleration scenario for the
flow \citep[cf.,][]{marscher1995}, and to test several alternative
acceleration scenarios including hydrodynamic acceleration
\citep{bodo1985}, acceleration by tangled magnetic field
\citep{heinz2000}, and magnetohydrodynamics acceleration
\citep{vlahakis2004}.

\begin{longtab}
\tiny
\begin{longtable}{ccccccccc}
\caption{\label{list of sources} List of Sources}\\
\hline\hline
\mc{Source (J2000)}   & Source (B1950) & Common Name & Epoch & $\alpha_\mathrm{2000}$ & $\delta_\mathrm{2000}$ &  $z$ & Type & $m_\mathrm{v}$  \\
\mc{(1)}        & (2) & (3) & (4) & (5) & (6) & (7) & (8) & (9) \\
\hline
\endfirsthead
\caption{List of Sources \textit{(continued)}}\\
\hline\hline
\mc{Source (J2000)}     & Source (B1950) & Common Name & Epoch & $\alpha_\mathrm{2000}$ & $\delta_\mathrm{2000}$ &  $z$ & Type & $m_\mathrm{v}$  \\
\mc{(1)}        & (2) & (3) & (4) & (5) & (6) & (7) & (8) & (9) \\
\hline
\endhead
\hline
\endfoot
J0013+4051 & 0010+405   & 4C 40.01     & C         & 00 13 31.130201 & +40 51 37.14403 & 0.2560 & G    & 17.9   \\
J0017+8135 & 0014+813   &      & B         & 00 17 08.474904 & +81 35 08.13656 & 3.3660 & Q    & 16.5     \\
J0030+7037 & 0027+703   &      & B         & 00 30 14.412959 & +70 37 40.06069 & …      & U    & 17.0                    \\
J0034+2754 & 0032+276   &      & C         & 00 34 43.486179 & +27 54 25.72112 & 2.9642 & G    & 18.0                 \\
J0044+6803 & 0041+677   &      & B         & 00 44 50.759603 & +68 03 02.68574 & …      & U    & …                     \\
J0046+2456 & 0043+246   &      & A         & 00 46 07.825730 & +24 56 32.52437 & 0.7467 & Q    & 17.1                   \\
J0057+3021 & 0055+300   &  NGC 315     & A         & 00 57 48.883342 & +30 21 08.81194 & 0.0165 & G    & 12.2                   \\
J0102+5824 & 0059+581   &  7C  0059+5808    & B         & 01 02 45.762380 & +58 24 11.13659 & 0.6440 & Q    & 17.6                    \\
\end{longtable}
\footnotesize{ {\bf Columns:} 1~--~IAU Source name (J2000); 2~--~IAU Source name (B1950); 3~--~common name;
4~--~observing epochs -- A: October 2010; B: May 2011 and C: October 2011; 5,6~--~source coordinates in J2000 epoch: right ascension and declination; 7~--~redshift;
8~--~optical class -- Q: quasar; B: BL Lac object; G: Galaxy;
U: Unidentified source; 9~--~optical V magnitude (information for Columns 7,8,9 obtained from the 
Simbad Astronomical Database \citep[ http://simbad.u-strasbg.fr/simbad;][]{wenger2000},
Sloan Digital Sky Survey (http://www.sdss.org/) and NASA/IPAC Extragalactic Database (https://ned.ipac.caltech.edu);  
(This table is also available in a machine-readable and Virtual Observatory (VO) forms in the online journal.)}
\end{longtab}

\begin{longtab}
\centering
\tiny
%\begin{landscape}
\begin{longtable}{cccccccccccccc}
\caption{\label{Image Parameters} Image Parameters}\\
\hline\hline

\mc{Source (J2000)}      & Obs & $S_\mathrm{86\,GHz}$ & $S_\mathrm{S}$ & $B_\mathrm{S}$ & $S_\mathrm{L}$  & $B_\mathrm{L}$  & $B_\mathrm{a}$ & $B_\mathrm{b}$  & $B_\mathrm{PA}$  & 
$S_\mathrm{t}$ & $S_\mathrm{p}$  & $\sigma$ & $\xi_\mathrm{r}$  \\
\mc{(1)}        &  (2) & (3) & (4) & (5) & (6) & (7) & (8) & (9)& (10)& (11)& (12)& (13)& (14) \\

\hline
\endfirsthead
\caption{Image Parameters \textit{continued}}\\
\hline\hline

\mc{Source (J2000)}   & Obs & $S_\mathrm{86\,GHz}$ & $S_\mathrm{S}$ & $B_\mathrm{S}$ & $S_\mathrm{L}$  & $B_\mathrm{L}$  & $B_\mathrm{a}$ & $B_\mathrm{b}$  & $B_\mathrm{PA}$  & 
$S_\mathrm{t}$ & $S_\mathrm{p}$  & $\sigma$ & $\xi_\mathrm{r}$  \\
\mc{(1)}        &  (2) & (3) & (4) & (5) & (6) & (7) & (8) & (9)& (10)& (11)& (12)& (13)& (14) \\

\hline
\endhead
\hline
\endfoot  
J0013+4051 & C         & 0.79                & 0.656  $\pm$  0.020  & 56       & 0.192  $\pm$  0.006 & 3136     & 159      & 35       & -16.8       & 436    & 317   & 8           & 1.43 \\
J0017+8135 & B         & 0.18                & 0.263  $\pm$  0.018  & 60       & 0.075  $\pm$  0.004 & 3062     & 74       & 35       & 62.1        & 143    & 81    & 2           & 1.36 \\
J0030+7037 & B         & 0.34                & 0.363  $\pm$  0.026  & 65       & 0.115  $\pm$  0.009 & 3070     & 76       & 36       & 66.5        & 155    & 92    & 3           & 1.34 \\
J0034+2754 & C         & 0.07                & 0.124  $\pm$  0.015  & 54       & 0.156  $\pm$  0.021 & 3086     & 262      & 36       & -14.2       & 159    & 60    & 2           & 1.04 \\
J0044+6803 & B         & 0.17                & 0.315  $\pm$  0.023  & 65       & 0.097  $\pm$  0.006 & 3034     & 75       & 36       & 70.4        & 133    & 99    & 3           & 1.15 \\
J0046+2456 & A         & 0.47                & 0.252  $\pm$  0.019  & 59       & 0.137  $\pm$  0.009 & 3060     & 267      & 37       & -15.0       & 235    & 160   & 6           & 1.17 \\
J0057+3021 & A         & 0.48                & 0.361  $\pm$  0.023  & 58       & 0.233  $\pm$  0.017 & 3060     & 308      & 37       & -14.5       & 306    & 226   & 8           & 1.07 \\
J0102+5824 & B         & 3.11                & 3.579  $\pm$  0.037  & 44       & 0.336  $\pm$  0.020 & 3050     & 59       & 43       & -19.5       & 2221   & 1075  & 19          & 1.58 \\
\end{longtable}
\footnotesize{ {\bf Columns:} (1)~--~IAU Source name (J2000);
  (2)~--~observing epochs -- A: October 2010; B: May 2011 and C:
  October 2011; (3)~--~total single dish flux density measured at
  86\,GHz obtained from the pointing and calibration scan measurements
  at Pico Veleta or Plateau de Bure [Jy]; (4),(6)~--~correlated flux densities [Jy]
  measured on projected baseline lengths listed in columns (5) and (7) [$\mathrm{M}\lambda$];
  (8)~--~major axis of the restoring beam [$\mu$as]; (9)~--~minor axis
  of the restoring beam [$\mu$as]; (10)~--~position angle of the major axis
  [degrees]; (11)~--~total clean flux density [mJy]; (12)~--~peak flux
  density [mJy/beam]; (13)~--~off-source r.m.s noise in the residual image
  [mJy/beam]; (14)~--~quality factor of the residual noise in the image.
  (This table is also available in a machine-readable and Virtual Observatory (VO) forms in the online journal.)}
%\end{landscape}
\end{longtab}

\begin{longtab}
\centering
\tiny
%\begin{landscape}
\begin{longtable}{ccccccccccc}
\caption{ \label{Model fit parameters} Model Fit Parameters}\\
\hline\hline
\mc{Source (J2000)}     & Obs &  Comp & $S_\mathrm{tot}$ &  $S_\mathrm{peak}$ & \textit{d} & \textit{r} & $\theta$ &  $T_\mathrm{b}$  &  $T_\mathrm{b,min}$   &  $T_\mathrm{b,lim}$  \\
\mc{(1)}        & (2) & (3) & (4) & (5) & (6) & (7) & (8) & (9) & (10) & (11) \\ 
\hline
\endfirsthead
\caption{Model Fit Parameters \textit{continued}}\\
\hline\hline
\mc{Source (J2000)}     & Obs &  Comp & $S_\mathrm{tot}$ &  $S_\mathrm{peak}$ & \textit{d} & \textit{r} & $\theta$ &  $T_\mathrm{b}$  &  $T_\mathrm{b,min}$   &  $T_\mathrm{b,lim}$  \\
\mc{(1)}        & (2) & (3) & (4) & (5) & (6) & (7) & (8) & (9) & (10) & (11) \\ 
\hline
\endhead
\hline
\endfoot
J0013+4051 & C & 1          & 288 $\pm$ 71     & 278 $\pm$ 49    & \textless14            & …                & …              & \textgreater28.642  & 4.46          & 9.69          \\
J0013+4051 & C & 2          & 194 $\pm$ 61     & 136 $\pm$ 35    & 34 $\pm$ 9                 & 41 $\pm$ 4           & -30.5$\pm$ 6.2     & 3.428 $\pm$ 1.647       & …             & …             \\
J0017+8135 & B & 1          & 99 $\pm$ 22      & 69 $\pm$ 12     & 30 $\pm$ 5                 & …                & …              & 8.115 $\pm$ 2.745       & 1.89          & 3             \\
J0017+8135 & B & 2          & 35 $\pm$ 18      & 11 $\pm$ 5      & 77 $\pm$ 38                & 431 $\pm$ 19         & -179.3$\pm$ 2.5    & 0.435 $\pm$ 0.377       & …             & …             \\
J0030+7037 & B & 1          & 121 $\pm$ 26     & 90 $\pm$ 16     & 26 $\pm$ 5                 & …                & …              & 2.947 $\pm$ 0.970       & 1.78          & 2.26          \\
J0030+7037 & B & 2          & 41 $\pm$ 45      & 4 $\pm$ 4       & 22 $\pm$ 24                & 53 $\pm$ 12          & 130.3$\pm$ 12.7    & 1.395 $\pm$ 2.651       & …             & …             \\
J0034+2754 & C & 1          & 72 $\pm$ 17      & 57 $\pm$ 10     & 26 $\pm$ 5                 & …                & …              & 6.803 $\pm$ 2.367       & 3.29          & 4.64          \\
J0034+2754 & C & 2          & 88 $\pm$ 36      & 13 $\pm$ 5      & 777 $\pm$ 316              & 758 $\pm$ 158        & -49.2$\pm$ 11.8    & 0.010 $\pm$ 0.007       & …             & …             \\
\end{longtable}
\footnotesize{ {\bf Columns:} (1)~--~IAU Source name(J2000);
  (2)~--~observing epochs -- A: October 2010; B: May 2011 and C:
  October 2011; (3)~--~I.D. number of Gaussian model fit component;
  (4)~--~total flux density of the component [mJy]; (5)~--~peak flux
  density of the component [mJy/beam]; (6)~--~component size
  [$\mu$as], with upper limits shown in italics;
  (7)~--~component's offset from the core [$\mu$as]; (8)~--~position angle of the offset [degrees]; (9)~--~
  brightness temperature obtained from the model fits [$ \times 10^{10}$\,K], with lower limits shown in italics; 
  (10)~--~visibility based estimate of the minimum brightness temperature [$ \times
  10^{10}$\,K]; (11)~--~visibility based estimate of the maximum
  resolved brightness temperature [$ \times 10^{10}$\,K].
  (This table is also available in a machine-readable and Virtual Observatory (VO) forms in the online journal.)}
%\end{landscape}
\end{longtab}
 
\section*{Acknowledgments}
\small
We thank the staff of the observatories participating in
the GMVA, the MPIfR Effelsberg 100 m telescope, the IRAM
Plateau de Bure Interferometer, the IRAM Pico Veleta 30 m telescope, the
Mets\"ahovi Radio Observatory, the Onsala Space Observatory,
and the VLBA. The VLBA is an instrument of the National Radio Astronomy Observatory,
which is a facility of the National Science Foundation operated
under cooperative agreement by Associated Universities, Inc.
This research has made use of the NASA/IPAC Extragalactic Database (NED) which is operated by the Jet Propulsion Laboratory, 
California Institute of Technology, under contract with the National Aeronautics and Space Administration.
This research has made use of the SIMBAD database, operated at CDS, Strasbourg, France
and also the Sloan Digital Sky Survey (SDSS). \par

Dhanya G. Nair is a member of the International Max Planck Research
School (IMPRS) for Astronomy and Astrophysics at the Universities of Bonn
and Cologne. Thanks to Biagina Boccardi, Jun Liu, Laura Vega García, Jae-Young Kim, 
Ioannis Myserlis, Vassilis Karamanavis, Jeff Hodgson, Shoko Koyama, Bindu Rani
and Karl M. Menten for their valuable suggestions and support in this research.
The author also thanks Walter Alef and Alessandra Bertarini for helping in the 
correlation of the 86\,GHz VLBI data used in this research. Thanks to 
Uwe Bach and Salvador Sánchez who have helped
in the observation and calibration at Effelsberg radio telescope and 
IRAM Pico Veleta radio telescope, respectively.
Sang-Sung Lee was supported by the National Research Foundation of Korea (NRF) grant funded by 
the Korea government (MSIP) (No. NRF-2016R1C1B2006697).
Yuri Y. Kovalev was supported in part by the government of the Russian Federation (agreement 05.Y09.21.0018) 
and by the Alexander von Humboldt Foundation.

\bibliographystyle{aa}
\bibliography{3mm}

\end{document}